\documentclass[preprint,authoryear, 13pt, a4paper]{elsarticle}
\usepackage{graphicx}
\usepackage{amssymb}
\usepackage{amsmath}
\usepackage{times,fancyhdr}
\usepackage{amsfonts}

\begin{document}

\begin{frontmatter}

\title{Molecular dynamics studies on the NMR and X-ray structures of rabbit prion proteins}
\author{Jiapu Zhang$^{1*}$, Yuanli Zhang$^2$}
\address{$^1$Graduate School of Sciences, Information Technology and Engineering, \& Centre of Informatics and Applied Optimization, The University of Ballarat,\\
 Mount Helen Campus, MT Helen, Victoria 3353, Australia;\\
$^2$School of Basic Medical Sciences, Taishan Medical University, Shandong 271000, China.\\  
$^*$Correspondence address: Tel: +61-3-5327 6335; Email: j.zhang@ballarat.edu.au
}

\begin{abstract}
Prion diseases, traditionally referred to as transmissible spongiform encephalopathies (TSEs), are invariably fatal and highly infectious neurodegenerative diseases that affect a wide variety of mammalian species, manifesting as scrapie in sheep and goats, bovine spongiform encephalopathy (BSE or mad-cow disease) in cattle, chronic wasting disease in deer and elk, and Creutzfeldt-Jokob diseases, Gerstmann-Strussler-Scheinker syndrome, fatal familial insomnia, and kulu in humans, etc. These neurodegenerative diseases are caused by the conversion from a soluble normal cellular prion protein (PrP$^{\text{C}}$) into insoluble abnormally folded infectious prions (PrP$^{\text{Sc}}$), and the conversion of {\it PrP$^{\text{C}}$ to PrP$^{\text{Sc}}$} is believed to involve conformational change from {\it a predominantly $\alpha$-helical protein to one rich in $\beta$-sheet structure}. Such a conformational change may be amenable to study by molecular dynamics (MD) techniques. For rabbits, classical studies show they have a low susceptibility to be infected by PrP$^{\text{Sc}}$, but recently it was reported that rabbit prions can be generated through saPMCA (serial automated Protein Misfolding Cyclic Amplification) in vitro and the rabbit prion is infectious and transmissible. In this paper, we first do a detailed survey on the research advances of rabbit prion protein (RaPrP) and then we perform MD simulations on the NMR and X-ray molecular structures of rabbit prion protein wild-type and mutants. The survey shows to us that rabbits were not challenged directly in vivo with other known prion strains and the saPMCA result did not pass the test of the known BSE strain of cattle. Thus, we might still look rabbits as a prion resistant species. MD results indicate that the three $\alpha$-helices of the wild-type are stable under the neutral pH environment (but under low pH environment the three $\alpha$-helices have been unfolded into $\beta$-sheets), and the three $\alpha$-helices of the mutants (I214V and S173N) are unfolded into rich $\beta$-sheet structures under the same pH environment. In addition, we found an interesting result that the salt bridges such as ASP201--ARG155, ASP177--ARG163 contribute greatly to the structural stability of RaPrP.
\end{abstract}


\end{frontmatter}

\section{Introduction}
\label{introduction}
Bovine Spongiform Encephalopathy (BSE) (``Mad cow disease") belongs to a contagious type of Transmissible Spongiform Encephalopathies (TSEs). Scientists believe it is caused by Prions (the misfolding prion proteins) but they may have not yet solved the riddle of ``mad cow disease". This is due to a prion is neither a virus, a bacteria nor any microorganism so the disease cannot be caused by the vigilance of the organism immune system and it can freely spread from one species to another species. The humans exists the susceptibility of TSEs. For example, the human version of "mad cow disease" named Creutzfeldt-Jakob Disease (CJD) and variant CJD (vCJD) just happen randomly through infections of transplanted tissue or blood transfusions or consumption of infected beef products. Cat, mink, deer, elk, moose, sheep, goat, nyala, oryx, greater kudu, ostrich and many other animals are also susceptible to TSEs. However, rabbits, horses and dogs seem to be unaffected by Prions (\cite{vorberg_etal2003, khan_etal2010, polymenidou_etal2008, zhang2011_horse, zhang2011_dog}). Scientists do not know the reason.

The prion protein is a naturally occurring protein in vivo. Its lesions in brain are not caused by the vigilance of the immune system. Recent studies have found that the lesions led astray as long as by contact with other normal prion proteins. The cells are arranged in accordance with the instruction of the gene and formed into proteins with different shapes and functions. But, like cardboard boxes, proteins need to be properly ``folded" in order to ensure their normal work. When proteins are folded into the wrong shape, they do not work. Under normal circumstances, the cells will supervise these misfolded proteins and automatically decompose them. However, the supervision mechanism is not with 100\% insurance. Scientists found that the rate of decomposition of prions is not quick enough and these prions accumulate and change the cellular metabolism and eventually kill the cell. This leads to the death of neurons in the brain. The dead neurons decompose and release more prion proteins into the biological mechanism to cause prion diseases. The infectious diseased prion is thought to be an abnormally folded isoform (PrP$^{\text{Sc}}$) of a host protein known as the prion protein (PrP$^{\text{C}}$). The conversion of PrP$^{\text{C}}$ to PrP$^{\text{Sc}}$ occurs post-translationally and involves conformational change from a predominantly $\alpha$-helical protein to one rich in $\beta$-sheet amyloid fibrils. Much remains to be understood about how the normal cellular isoform of the prion protein PrP$^{\text{C}}$ undergoes structural changes to become the disease associated amyloid fibril form PrP$^{\text{Sc}}$. The ``structural conformational" changes of PrP$^{\text{Sc}}$ from PrP$^{\text{C}}$ are just very proper to be studied by MD techniques. Classical studies have showed that rabbits have low susceptibility to be infected by PrP$^{\text{Sc}}$ (\cite{vorberg_etal2003, khan_etal2010, barlow_etal1976, fernandez-funez_etal2009, korth_etal1997, courageot_etal2008, vilette_etal2001, nisbet_etal2010, wen_etal2010a, wen_etal2010b, zhou_etal2011, ma_etal2012}). However, recently there is one exception saying that (i) rabbit prion can be produced through saPMCA ({\it s}erial {\it a}utomated Protein Misfolding Cyclic Amplification) in vitro (\cite{chianini_etal2012, fernandez-borges_etal2012}) (though not by challenging rabbits directly in vivo with other known prion strains), and (ii) the rabbit prion generated is infectious and transmissible (\cite{chianini_etal2012}). RaPrP had already had NMR (\cite{li_etal2007}) and X-ray (\cite{khan_etal2010}) structures in Protein Data Bank (www.rcsb.org) with PDB entries 2FJ3, 3O79 respectively. In 2010 and 2013, the NMR and X-ray structures of some RaPrP mutants (with PDB entries 2JOH, 2JOM, 4HLS, 4HMR, 4HMM) were also already released into PDB bank (\cite{wen_etal2010b, sweeting_etal2013}). This paper will study the NMR and X-ray structures of RaPrP and its I214V and S173N mutants by MD techniques, in order to explain the specific mechanism about rabbit PrP$^{\text{C}}$ (RaPrP$^{\text{C}}$) and the conversion of PrP$^{\text{C}}\rightarrow$PrP$^{\text{Sc}}$ of rabbits. Surely, the studies should provide valuable knowledge about the rules governing the PrP$^{\text{C}}\rightarrow$PrP$^{\text{Sc}}$ conversion, which can provide some ideas for designing novel therapeutic approaches that block the conversion and disease propagation.

In the MD studies, we are not only doing simulations under neutral pH environment, but also under low pH environment, because recent studies indicate that the low pH solution system is an ideal trigger of PrP$^{\text{C}}$ to PrP$^{\text{Sc}}$ conversion (\cite{gerber_etal2008, bjorndahl_etal2011, zhouh2013}). 

Through some preliminary analyses of the MD simulation results, we have found that salt bridge is a clear factor of the structural stability of RaPrP. This agrees with a recent finding on salt bridges about PrP$^{\text{C}}$ (\cite{zhouh2013}). The salt bridges of the native prion protein (PrP$^{\text{C}}$) have been calculated and analyzed by quantum chemical calculations by Ishikawa et al. (2010) (\cite{ishikawa_etal2010}), and the detailed biophysical characteristics and NMR studies of PrP$^{\text{C}}\rightarrow$PrP$^{\text{Sc}}$ conversion process have also been reported by Bjorndahl et al. (2011) (\cite{bjorndahl_etal2011}). Recently, incorporating Chou's wenxiang diagrams (\cite{chou_etal1997, chou_etal2011, zhou2011a, zhou2011b, zhouh2013}) further summarized their research based on the NMR, CD (circular diagram) spectra and DLS (Dynamic light scattering) data at the low pH environment, and found that some salt bridges and the hydrophobic interactions in the three helices of the prion proteins can affect the helical structural stability. These studies have provided the insight into the prion misfolding mechanism.

The rest of this paper is organized as follows. Firstly, we will review previous research results on rabbit prion protein listed in the PubMed of NCBI. Secondly, we will present the MD simulation materials and methods for NMR and X-ray structures of RaPrP$^{\text{C}}$ wild-type and mutants. Thirdly, we will give analyses of MD simulation results and discussions. Lastly, concluding remarks on RaPrP are summarized.

\section{A Detailed Review on Rabbit PrP}
The symptoms of TSEs were first described for sheep in 1730 and called ``scrapie" in England, ``vertige" in France and ``Traberkrankheit" in Germany (\cite{verdier2012}). Now we know that many species such as sheep, goats, mice, humans, chimpanzees, hamsters, cattle, elks, deers, minks, cats, chicken, pigs, turtles, etc are susceptible to TSEs. But many laboratory experiments show that rabbits, horses and dogs seem to be the resistant (or at least the low-susceptibility-rate) species to TSEs. However, recently Chianini {\it et al.} (2012) reported ``rabbits are not resistant to prion infection" (\cite{chianini_etal2012}). Thus, at this moment it is very necessary for us to give a detailed review some laboratory works (from Year 1976 to Year 2013) on RaPrP:
\begin{itemize}
\item In 1976, Barlow and Rennie (1976) made many attempts to infect rabbits with the ME7 scrapie (where scrapie is a prion disease in sheep and goats) and other known prion strains but all failed at last (\cite{barlow_etal1976}).
\item In 1984$\sim$1985, some antibodies to the scrapie protein were reported by Prusiner's research group (\cite{bendheim_etal1984, barry_etal1985}. The antibodies to the scrapie agent were produced after immunization of rabbits with either scrapie prions or the prion protein PrP(27--30). The monospecificity of the rabbit antiserum raised against PrP(27--30) was established by its reactivity after affinity purification, and the rabbit antiserum to PrP(27--30) was successfully produced. In 1985, the characterization of antisera raised in rabbits, against scrapie-associated prion diseases and the prion human CJD, was studied in (\cite{bode_etal1985}). 
\item In 1986, Takahashi et al. (1986) reported, rabbits immunized with the fraction P4 containing scrapie infectivity prepared from mouse brains raised antibodies against three major polypeptides of (\cite{takahashi_etal1986}). Cho (1986) reported that the antibody to scrapie-associated fibril protein roduced in a rabbit identifies a cellular antigen (\cite{cho1986}). Barry et al. (1986) reported that rabbit antisera to a synthetic peptide PrP-P1 constructed based on PrP(27--30) were found by immunoblotting to react with PrP(27--30) and its precursor PrP$^\text{Sc}$(33--35), as well as with a related protease-sensitive cellular homologue PrP$^\text{C}$(33--35), this means scrapie (PrP$^\text{Sc}$) and cellular (PrP$^\text{C}$) prion proteins share polypeptide epitopes (\cite{barry_etal1986}). An enzyme-linked immunosorbent assay showed that rabbit antiserum to PrP(27--30) was more reactive with PrP(27--30) than with PrP-P1; conversely, antiserum to PrP-P1 was more reactive with the peptide than with the prion proteins (\cite{barry_etal1986}).  Shinagawa et al. (1986) reported ``immunization of a rabbit with the (synthetic) peptide conjugated with ovalbumin induced specific antibodies" corresponding to the N-terminal region of the scrapie prion protein (PrP$^\text{Sc}$) (\cite{shinagawa_etal1986}). ``Rabbit antisera were raised to SAFs (scrapie-associated fibrils) isolated from mice infected with the ME7 scrapie strain and to SAFs isolated from hamsters infected with the 263K scrapie strain" (\cite{kascsak_etal1986, merz_etal1987}). Robakis et al. (1986) clearly pointed out that rabbit brain is resistant to scrapie infection (\cite{robakis_etal1986}) in 1986. 
\item In 1987, there were several reports on rabbits. Bockman et al. (1987) identified by immunoblotting human \& mouse CJD prion proteins (HuPrP$^\text{Sc}$ and MoPrP$^\text{Sc}$) using rabbit antisera raised against hamster scrapie prion proteins (HaPrP$^\text{Sc}$) (\cite{bockman_etal1987}). Wade et al. (1987) found a 45 kD protein in scrapie-infected hamster brain has a signal to inoculate rabbits (\cite{wade_etal1987}). Kascsak et al. (1987) reported ``MAb (monoclonal antibody) 263K 3F4 (that was derived from a mouse immunized with hamster 263K PrP$^\text{Sc}$ reacted with hamster but not mouse PrP$^\text{Sc}$)  recognized normal host protein of 33 to 35 kilodaltons in brain tissue from hamsters and humans but not from bovine, mouse, rat, sheep, or rabbit brains" (\cite{kascsak_etal1987}). Wiley et al. (1987) used rabbit monospecific antisera raised against synthetic peptides corresponding to the N-terminal 13 or 15 amino acids of PrP(27--30) and rabbit antisera raised against infectious prions or PrP(27--30) purified from scrapie-infected hamster brains to immunostaine glutaraldehyde-perfused hamster brains (\cite{wiley_etal1987}). Hay et al. (1987) found the evidence for a secretory form of the cellular prion protein (PrP$^\text{C}$) ``cell-free translation studies in rabbit reticulocyte lysates supplemented with microsomal membranes gave results: while one form of HaPrP (hamster brain prion protein) was found as an integral membrane protein spanning the membrane at least twice, another form of HaPrP was found to be completely translocated to the microsomal membrane vesicle lumen" (\cite{hay_etal1987}).
\item In 1988, Caughey et al. (1988) detected the immunoprecipitation of PrP synthesis using a rabbit antibody specific for a 15 amino acid PrP peptide and concluded that ``either PrP is not the transmissible agent of scrapie or the PrP is not processed appropriately in this cell system to create the infectious agent" (\cite{caughey_etal1988}). Barry et al. (1988) undertook ELISA (enzyme-linked immunosorbent assay) and immunoblotting studies with rabbit antisera raised against three synthetic PrP peptides of PrP(27--30), PrP$^\text{Sc}$, and PrP$^\text{C}$ and concluded that the three proteins are encoded by the same chromosomal gene (\cite{barry_etal1988}). Baron et al. (1998) found ``polyclonal rabbit antiserum to SAF protein was reacted with brain sections from scrapie-infected mice, two familial cases of transmissible dementia, and three cases of Alzheimer's disease (AD)" and ``evidence of the similarity of SAF protein to PrP(27--30)" (\cite{baron_etal1988}). Gabizon et al. (1988) found ``polyclonal RaPrP antiserum raised against NaDodSO4/PAGE-purified scrapie prion protein of 27--30 kDa reduced scrapie infectivity dispersed into detergent-lipid-protein complexes" (\cite{gabizon_etal1988}). Roberts et al. (1998), ``using monoclonal antibodies to a synthetic peptide corresponding to a portion of beta-protein and rabbit antiserum to hamster scrapie PrP(27--30), examined in situ amyloid plaques on sections from cases of neurodegenerative diseases" and their ``results emphasize the need for classification of CNS (central nervous system) amyloids based on the macromolecular components comprising these pathologic polymers" (\cite{roberts_etal1988}).
\item In 1989, Gabizon et al. (1989) reported that ``polyclonal rabbit PrP antiserum raised against sodium dodecyl sulfate-polyacrylamide gel electrophoresis (SDS-PAGE)-purified PrP(27--30) reduced scrapie infectivity dispersed into DLPC (detergent-lipid-protein complexes)" (\cite{gabizon_etal1989}). ``Kuru plaque is a pathognomonic feature in the brain of patients with CJD and in the brain of CJD-infected mice" (\cite{kitamoto_etal1989}), Kitamoto et al. (1989) reported ``kuru plaques from CJD-infected mice were immunolabeled with rabbit anti-murine prion protein (PrP) absorbed with human PrP, but not so with mouse anti-human PrP" (\cite{kitamoto_etal1989}). Farquhar et al. (1989) ``Two polyclonal antisera were raised in rabbits against the scrapie-associated fibril protein (PrP) prepared from sheep and mice" (\cite{farquhar_etal1989}).
\item In 1990, Yost et al. (1990) reported that in the rabbit reticulocyte lysate system, an unusual topogenic sequence in the prion protein fails to cause stop transfer (the polypeptide chain across the membrane of the endoplasmic reticulum) of most nascent chains (\cite{yost_etal1990}). Lopez et al. (1990) reported a completely translocated (secretory) topology form of the major product synthesized in rabbit reticulocyte lysates (RRL) (\cite{lopez_etal1990}).
\item In 1991, Di Martino et al. (1991) reported the characterization of two polyclonal antibodies which were raised by immunizing rabbits with two non carrier-linked synthetic peptides whose amino acid sequences corresponded to codons 89--107 (peptide P1) and 219--233 (peptide P2) of the translated cDNA sequence of murine PrP protein (\cite{dimartino_etal1991}). Ikegami et al. (1991) detected the scrapie-associated fibrillar protein in the lymphoreticular organs of sheep by means of a rabbit-anti-sheep PrP (the scrapie-associated fibrillar protein) polyclonal antibody by Western blot analysis (\cite{ikegami_etal1991}).
\item In 1992, Hashimoto et al. (1992) did immunohistochemical study of kuru plaques using antibodies against synthetic prion protein peptides and used two synthetic peptides to immunize rabbits and produce antisera (anti-N and anti-M) (\cite{hashimoto_etal1992}). Kirkwood et al. (1992) using rabbit antiserum raised against mouse PrP detected an abnormal PrP (prion protein) from the brains of domestic cattle with spongiform encephalopathy (SE) (\cite{kirkwood_etal1992}).
\item In 1993, the Western blot analysis was performed with rabbit serum against the sheep SAF (\cite{onodera_etal1993}). To determine if amyloid deposits be visualized by immunocytochemical techniques, Guiroy et al. (1993) used a rabbit antiserum directed against scrapie amyloid PrP(27--30) to stain formalin-fixed, formic acid-treated brain tissue sections from several animal species with natural and experimental transmissible mink encephalopathy (TME) (\cite{guiroy_etal1993}). Groschup and Pfaff (1993) reported that ``rabbit antisera to synthetic peptides representing amino acid sequence 108 to 123 of PrP of cattle, sheep and mice reacted strongly with modified PrP of the homologous host but not, or only poorly, with PrP of heterogeneous origin" (\cite{groschupp1993}). Miller et al. (1993) did the immunohistochemical detection of prion protein in sheep with scrapie using a primary antibody obtained from a rabbit immunized to PrP$^\text{Sc}$ extracted from brains of mice with experimentally induced scrapie (\cite{miller_etal1993}).
\item In 1994, Groschup et al. (1994) investigated with eight different anti-peptide sera raised in rabbits against various synthetic peptides representing segments of the amino acid (aa) sequence 101--122 of ovine, bovine, murine and hamster PrP, and found that ``the region close to the actual or putative proteinase K cleavage sites of PrP seems to exhibit high structural variability among mammalian species" (\cite{groschup_etal1994}).  Xi et al. (1994) detected the proteinase-resistant protein (PrP) in small brain tissue samples from CJD patients using rabbit polyclonal antibody against hamster PrP(27--30) (\cite{xi_etal1994}). Schmerr et al. (1994) used a fluorescein-labeled goat anti-rabbit immunoglobulin as an antibody and used rabbit antiserum for immunoblot analysis, and PrP$^\text{Sc}$ was solubilized and reacted with a rabbit antiserum specific for a peptide of the prion protein (\cite{schmerr_etal1994}).  
\item In 1995$\sim$1996, Yokoyama et al. did some works. In 1995, Yokoyama et al. (1995) used antisera raised in rabbits against three peptides PrP 150--159, PrP 165--174, and PrP 213--226 of mouse prion and concluded that rabbit antiserum against the MAP (multiple antigenic peptide) representing amino acid sequence 213--226 of mouse PrP is useful as a diagnostic tool for prion disease of animals (\cite{yokoyama_etal1995}). In 1996, Yokoyama et al. (1996) detected species specific epitopes of mouse and hamster prion proteins by anti-peptide antibodies, where the antisera were produced in rabbits (\cite{yokoyama_etal1996}). 
\item In 1997, Madec et al. (1997) undertook Western blot analyses using rabbit antiserum that recognized both normal and pathologic sheep prion proteins to study the biochemical properties of PrPSc in natural sheep scrapie (\cite{madec_etal1997}).  Loftus and Rogers (1997) cloned RaPrP open reading frame (ORF) and characterised rabbits as a species with apparent resistance to infection by prions (\cite{loftusr1997}). Groschup et al. (1997) raised antisera in rabbits and chicken against sixteen synthetic peptides which represent the complete amino acid sequence of ovine PrP to generate antibodies to further regions of PrP, in order to immunochemical diagnosis and pathogenetic studies on prion diseases (\cite{groschup_etal1997}). In (\cite{komar_etal1997}),``the Ure2p yeast prion-like protein was translated in vitro in the presence of labeled [35S]methionine in either rabbit reticulocyte lysate (RRL) or wheat germ extract (WGE) cell-free systems". In 1997, Korth et al. (1997) found that RaPrP was not recognized by a conformational antibody specific for PrPSc-like structures (\cite{korth_etal1997}). 
\item In 1999, Takahashi et al. (1999) immunized rabbits with four synthetic peptides and compared the immunoreactivity of antibodies to bovine prion proteins (bovine-PrPs) from various species by immunoblotting and immunohistochemistry (\cite{takahashi_etal1999}) and they identified two regions in bovine-PrP which appear suitable for raising antibodies that detect various kinds of PrPs, and one region (Ab103--121) which appears suitable for raising antibodies that detect several species of PrP (\cite{takahashi_etal1999}).
\item In 2000, Garssen et al. (2000) did applicability of three anti-PrP peptide sera including staining of tonsils and brainstem of sheep with scrapie (\cite{garssen_etal2000}). ``The three rabbit antibodies (R521, R505, R524) were produced, and raised to synthetic peptides corresponding to residues 94--105, 100--111, and 223--234, respectively, of the sheep prion protein" (\cite{garssen_etal2000}). ``The usefulness of all three anti-peptide sera in the immunohistochemical detection of PrP$^\text{Sc}$ in brain stem and tonsils of scrapie-affected sheep was demonstrated and compared with an established rabbit anti-PrP serum" (\cite{garssen_etal2000}). Zhao et al. (2000) using prokaryotic expressed GST-PrP fusion protein as antigen, found that ``rabbits were immunized subcutaneously" (\cite{zhao_etal2000}).
\item In 2001, Kelker et al. (2001) showed that ``combination of authentic rabbit muscle GAPDH (glyceraldehyde-3-phosphate dehydrogenase) with tNOX (a cell surface NADH oxidase of cancer cells) renders the GAPDH resistant to proteinase K digestion" (\cite{kelker_etal2001}). Vol'pina et al. (2001) reported ``rabbits were immunized with either free peptides or peptide-protein conjugates to result in sera with a high level of antipeptide antibodies" to the BSE prion disease (\cite{volpina_etal2001}). Bencsik et al. (2001) identified prion protein PrP ``using either RB1 rabbit antiserum or 4F2 monoclonal antibody directed against AA 108--123 portion of the bovine and AA 79--92 of human prion protein respectively" and ``showed the close vicinity of these PrP expressing cells with noradrenergic fibers" (\cite{bencsik_etal2001}).  In (\cite{li_etal2001}), ``the rabbits were immuned with bovine prion protein (BoPrP$^\text{C}$) which was expressed in E. coli and anti-PrP$^\text{C}$ antibody (T1) was obtained", and Li et al. (2001) could detect BSE and scrapie with T1 antibody (\cite{li_etal2001}).
\item In 2002, Laude et al. (2002) reported ``In one otherwise refractory rabbit epithelial cell line, a regulable expression of ovine PrP was achieved and found to enable an efficient replication of the scrapie agent in inoculated cultures" (\cite{laude_etal2002}). Takekida et al. (2002) established a competitive ELISA to detect prion protein in food products using rabbit polyclonal antibodies that were raised against bovine prion peptides (\cite{takekida_etal2002}).
\item In 2003, Vorberg et al. (2003) found multiple amino acid residues (such as GLY99, MET108, SER173, ILE214) within the 	RaPrP inhibit formation of its abnormal isoform (\cite{vorberg_etal2003}). The authors made some substitutions of mouse PrP amino acid sequence by rabbit PrP amino acid sequence and found (i) at the N-terminal region (residues 1--111) the PrPSc formation is totally prevented, (ii) at the central region (residues 112--177), the constructed PrP failed to be converted to protease resistance, (iii) at the C-terminal region (residues 178--254) the formation of PrP$^\text{Sc}$ is drastically decreased but is not abolished completely (\cite{vorberg_etal2003}). Thus, rabbit cells are negatively affected by the formation of PrP$^\text{Sc}$. Jackman and Schmerr (2003) synthesized fluorescent peptides from the prion protein and produced the corresponding antibodies in rabbits against these peptides, and at last detected abnormal prion protein in a tissue sample (\cite{jackmans2003}). Gilch et al. (2003) reported ``treatment of prion-infected mouse cells with polyclonal anti-PrP antibodies generated in rabbit or auto-antibodies produced in mice significantly inhibited endogenous PrPSc synthesis" and found ``immune responses against different epitopes when comparing antibodies induced in rabbits and PrP wild-type mice" (\cite{gilch_etal2003}). 
\item In 2004, Brun et al. (2004) reported the development and further characterisation of a novel PrP-specific monoclonal antibody 2A11, which reacts with PrPC from a variety of species including rabbit (\cite{brun_etal2004}). Sachsamanoglou et al. (2004) described ``the quality of a rabbit polyclonal antiserum (Sal1) that was raised against mature human recombinant prion protein (rHuPrP)" (\cite{sachsamanoglou_etal2004}). Senator et al. (2004) investigated ``the effects of cellular prion protein (PrP$^\text{C}$) overexpression on paraquat-induced toxicity by using an established model system, rabbit kidney epithelial A74 cells, which express a doxycycline-inducible murine PrPC gene" (\cite{senator_etal2004}).
\item In 2005, Golanska et al. (2005) used 2 different anti-14-3-3 antibodies: rabbit polyclonal and mouse monoclonal antibodies to analyze the 14-3-3 protein in the cerebrospinal fluid in CJD (\cite{golanska_etal2005}).
\item In 2006, Dupiereux et al. (2006) investigated the effect of PrP(106-126) peptide on an established non neuronal model, rabbit kidney epithelial A74 cells that express a doxycycline-inducible murine PrPC gene (\cite{dupiereux_etal2006}). Biswas et al. (2006) reported ``a rabbit polyclonal anti-serum raised against dimeric MuPrP (murine prion protein) cross-reacted with p46 (a 46 kDa species) and localized the signal within the Golgi apparatus" (\cite{biswas_etal2006}). Gao et al. (2006) reported recombinant neural protein PrP can bind with both recombinant and native apolipoprotein E (ApoE) in vitro, where the ``ApoE-specific antiserum was prepared by immunizing rabbits with the purified ApoE3" (\cite{gao_etal2006}). Kocisko and Caughey (2006) reported ``rabbit epithelial cells that produce sheep prion protein in the presence of doxycycline (Rov9) have been infected with sheep scrapie" (\cite{kociskoc2006}). Xiao et al. (2006) used the method ``two male rabbits were immunized for 4 times with the purified protein, and the antiserum against NSE protein was collected and evaluated by ELISA, Western blotting and immunohistochemistry"  and concluded ``high expression of HuNSE (human neuron-specific enolase) is obtained in E. coli and the prepared antiserum against HuNSE can be used potentially for diagnosis of prion-associated diseases and other nervous degeneration diseases" (\cite{xiao_etal2006}). 
\item In 2007, Oboznaya et al. (2007) reported ``antibodies to a nonconjugated prion protein peptide 95--123 interfere with PrP$^\text{Sc}$ propagation in prion-infected cells", where ``rabbits were immunized with free nonconjugated peptides" (\cite{oboznaya_etal2007}). Handisurya et al. (2007) reported ``Immunization with PrP-virus-like particles induced high-titer antibodies to PrP in rabbit and in rat, without inducing overt adverse effects. As determined by peptide-specific ELISA, rabbit immune sera recognized the inserted murine/rat epitope and also cross-reacted with the homologous rabbit/human epitope differing in one amino acid residue. Rabbit anti-PrP serum contained high-affinity antibody that inhibited de novo synthesis of PrP$^\text{Sc}$ in prion-infected cells" (\cite{handisurya_etal2007}). Bastian et al. (2007) did an experiment ``spiroplasma mirum, a rabbit tick isolate that had previously been shown to experimentally induce spongiform encephalopathy in rodents, was inoculated intracranially (IC) into ruminants" and at last concluded ``Spiroplasma spp. from TSE brains or ticks induce spongiform encephalopathy in ruminants" (\cite{bastian_etal2007}).  Dong et al. (2007) did the interaction analysis between various PrP fusion proteins and the tubulin in vitro, where the native tubulin was extracted from rabbit brain tissues (\cite{dong_etal2007}).
\item In 2008, it was said that ``ovine prion protein renders rabbit epithelial RK13 cells permissive to the multiplication of ovine prions, thus providing evidence that species barriers can be crossed in cultured cells through the expression of a relevant ovine PrP$^\text{C}$" (\cite{courageot_etal2008}). Sakudo et al. (2008) ``developed a mammalian expression system for a truncated soluble form of human prion protein with the native signal peptide but without a glycosylphosphatidylinositol (GPI)-anchor site, driven by the peptide chain elongation factor 1alpha promoter in stably transfected rabbit-kidney epithelial RK13 cells, to investigate the SOD (superoxide dismutase) activity of mammalian prion protein" and concluded ``GPI-anchorless human prion protein is secreted and glycosylated but lacks superoxide dismutase activity" (\cite{sakudo_etal2008}). Shin et al. (2008) cloned a prion protein (PrP) Glu218Lys gene from Korean bovine (Bos taurus coreanae) and raised the production of rabbit anti-bovine PrP antibody (\cite{shin_etal2008}). Lawson et al. (2008) reported ``rabbit kidney epithelial cells (RK13) are permissive to infection with prions from a variety of species upon expression of cognate PrP transgenes" (\cite{lawson_etal2008}).
\item In 2009, Hanoux et al. (2009) reported ``when injected into rabbits, (a synthetic peptide) CDR3L generated anti-SAF61 anti-Id polyclonal antibodies that exclusively recognized SAF61 mAb but were unable to compete with hPrP for antibody binding" (\cite{hanoux_etal2009}). Tang et al. (2009) reported fibrinogen, one of the most abundant extracellular proteins, has chaperone-like activity: it maintains thermal-denatured luciferase in a refolding competent state allowing luciferase to be refolded in cooperation with rabbit reticulocyte lysate, and it also inhibits fibril formation of yeast prion protein Sup35 (NM) (\cite{tang_etal2009}). Differed from the reaction with N-terminal proline/glycine-rich repeats recognizing rabbit polyclonal antibody, seven monoclonal antibodies (mAbs) against chicken cellular prion protein (ChPrP$^\text{C}$) were obtained (\cite{ishiguro_etal2009}). Fernandez-Funez et al. (2009) found RaPrP does not induce neurodegeneration in the brains of transgenic flies (\cite{fernandez-funez_etal2009}).
\item In 2010, Nisbet et al. (2010) created a mutant mouse PrP model containing RaPrP specific amino acids at the GPI anchor site and found that the GPI anchor attachment site ($\omega$ site) controls the ability of PrP$^\text{C} \rightarrow$PrP$^\text{Sc}$ and the residues at $\omega$ and $\omega$+1 of PrP are important modulators of this pathogenic process (\cite{nisbet_etal2010}). Nisbet et al. (2010) recognized that ``rabbits are one of a small number of mammalian species reported to be resistant to prion infection" (\cite{nisbet_etal2010}). Wen et al. (2010) using multidimensional heteronuclear NMR techniques reported that the I214V and S173N substitutions result in distinct structural changes for RaPrPC (\cite{wen_etal2010a, wen_etal2010b}) and concluded that the highly ordered $\beta$2-$\alpha$2 loop may contribute to the local as well as global stability of the RaPrP protein (\cite{wen_etal2010b}). Wen et al. (2010) also recognized ``rabbits are one of the few mammalian species that appear to be resistant to TSEs due to the structural characteristics of RaPrPC itself" (\cite{wen_etal2010b}). Fernandez-Funez et al. (2010) showed that ``RaPrP does not induce spongiform degeneration and does not convert into scrapie-like conformers" (\cite{fernandez-funez_etal2010}). Bitel et al. (2010) examined ``changes in muscle tissue in a classic model of diabetes and hyperglycemia in rabbits to determine if similar dysregulation of Alzheimer A$\beta$ peptides, the prion protein (PrP), and superoxide dismutase 1 (SOD1), as well as nitric oxide synthases is produced in muscle in diabetic animals" (\cite{bitel_etal2010}). Khan et al. (2010) found the propensity to form $\beta$-state (the $\beta$-sheet-rich structure) is greatest for hamster PrP, less for mouse PrP, but least for the PrP of rabbits, horses and dogs under different conditions and using two-wavelength CD (Circular Dichroism) method they also found a key hydrophobic staple-like helix-capping motif keeping the stability of RaPrP's X-ray crystallographic molecular structure (\cite{khan_etal2010}). 
\item In 2011, Zocche et al. (2011) used the methods ``rabbit aortic smooth muscle cells were challenged for 4, 8 and 18 hours, with angiotensin-II, tunicamycin and 7-ketocholesterol, and rabbit aortic arteries were subjected to injury by balloon catheter", and got the results ``the PrP$^\text{C}$ mRNA expression in rabbit aortic artery fragments, subjected to balloon catheter injury, showed a pronounced increase immediately after overdistension" (\cite{zocche_etal2011}). Mays et al. (2011) reported ``PrP$^\text{Sc}$ was efficiently amplified with lysate of rabbit kidney epithelial RK13 cells stably transfected with the mouse or Syrian hamster PrP gene" (\cite{mays_etal2011}). Julien et al. (2011) reported the different overall sensitivities toward NMR urea denaturation with stabilities in the order hamster = mouse $<$ rabbit $<$ bovine protein, and they also investigated the effect of the S174N mutation in rabbit PrP$^\text{C}$ (\cite{julien_etal2011}). Zhou et al. (2011) found that the crowded physiological agents Ficoll 70 and dextran 70 have effects significantly inhibiting fibrillation of RaPrP (\cite{zhou_etal2011, ma_etal2012}). Fernandez-Funez et al. (2011) also acknowledged that ``classic studies showing the different susceptibility to prion disease in mammals have recently found support in structural and transgenic studies with PrP from susceptible (mouse, hamster) and resistant (rabbit, horse, dog) animals" (\cite{fernandez-funez_etal2011}). 
\item In 2012, Chianini et al. (2012) generated rabbit PrP$^\text{Sc}$ in vitro subjecting unseeded normal rabbit brain homogenate to saPMCA and found the rabbit PrP$^\text{Sc}$ generated in vitro is infectious and transmissible (\cite{chianini_etal2012}) and they declared ``rabbits are not resistant to prion infection" (\cite{chianini_etal2012}). Kim et al. (2012) reported ``elk prion protein (ElkPrP$^\text{C}$) has been confirmed to be capable of rendering rabbit epithelial RK13 cells permissive to temporal infection by chronic wasting disease (CWD) prions." (\cite{kim_etal2012}). Fernández-Borges et al. (2012) reported the results of (\cite{chianini_etal2012}) and pointed out it is not reasonable to attribute species-specific prion disease resistance based purely on the absence of natural cases and incomplete in vivo challenges; the concept of species resistance to prion disease should be re-evaluated using the new powerful tools available in modern prion laboratories, whether any other species could be at risk (\cite{fernandez-borges_etal2012}).
\item In 2013, Vidal et al. (2013) studied the saPMCA and reported that rabbits are an apparently resistant species to the original classical cattle BSE prion (\cite{vidal_etal2013}). Wang et al. (2013) reported rabbits are ``insensitivity to prion diseases" (\cite{wangz_etal2013}). Wang et al. (2013) aimed to investigate ``potential mechanisms contributing to prion resistance/susceptibility by using the rabbit, a species unsusceptible to prion infection, as a model" (\cite{wangy_etal2013}) and investigated ``the expression level and distribution of LRP/LR (laminin receptor precursor/laminin receptor) in rabbit tissues by real-time polymerase chain reaction and by immunochemical analysis with a monoclonal anti-67 kDa LR antibody" (\cite{wangy_etal2013}) and at last their findings confirmed the prion resistance in rabbits (\cite{wangy_etal2013}). Sweeting et al. (2013) produced X-ray structures of mutants in the $\beta$2-$\alpha$2 loop and reported that the helix-capping motif in the $\beta$2-$\alpha$2 loop modulates $\beta$-state misfolding in RaPrP, and still acknowledged ``rabbit PrP, a resistant species" (\cite{sweeting_etal2013}). Friedman-Levi et al. (2013) reported ``pAb RTC and EP1802Y (EP), a rabbit $\alpha$ PrP mAb directed against the CITQYER ESQAYYQRGS sequence present at the C-terminal part of human PrP, just before the PrP GPI anchor" (\cite{friedman-levi_etal2013}). Timmes et al. (2013) used R20 as the rabbit polyclonal anti-PrP antibody (\cite{timmes_etal2013}).
\end{itemize}
\noindent Throughout the above review on the research advances in RaPrP, we noticed that the rabbit prion in (\cite{chianini_etal2012, fernandez-borges_etal2012}) was just produced through saPMCA in vitro not by challenging rabbits directly in vivo with other known prion strains, and the saPMCA result of (\cite{chianini_etal2012, fernandez-borges_etal2012}) was refused by the test of cattle BSE (\cite{vidal_etal2013}). All other RaPrP research results generally agree with each other to look rabbits as a resistant species to prion diseases.

\section{Materials and Methods}
Many marvelous biological functions in proteins and DNA and their profound dynamic mechanisms, such as cooperative effects (\cite{chou1989}), allosteric transition (\cite{chou1987}), DNA internal motion (\cite{zhou1989}), intercalation of drugs into DNA (\cite{choum1988}), and assembly of microtubules (\cite{chou_etal1994}), can be revealed by studying their internal motions as summarized in a comprehensive review (\cite{chou1988}). Likewise, to really understand the action mechanism of prion protein, we should consider not only the static structures concerned but also the dynamical information obtained by simulating their internal motions or dynamic process. To realize this, the MD simulation is one of the feasible tools.

The MD simulations of this paper are the continuation and extension of the ones of (\cite{zhang2010}). Zhang (2010) carried out (i) 15 ns of production phase of MD simulations and (ii) the heating phase of MD simulations is starting from 100 K with one set of initial velocity (denoted as seed2) (\cite{zhang2010}). This paper continued to finish another 15 ns of production phase of MD simulations for seed2, and carried out other two sets of initial velocities for heating from 100 K (denoted as seed1 and seed3) of MD simulations with 30 ns of production phases. 

We furthermore extended the MD simulation of (\cite{zhang2010}). Under the same simulation conditions, we also did the MD simulations with the three seeds at room temperature (300 K) for X-ray structures (PDB entries 3O79 (\cite{khan_etal2010}),  4HMM (\cite{sweeting_etal2013})) and the NMR structures (PDB entries 2FJ3, 2JOH, 2JOM), in order to further confirm the MD results of NMR structures. For the three seeds at room temperature and at 350 K, the MD simulations are also done for human and mouse prion proteins (PDB entries 1QLX (\cite{zahn_etal2000}), 1AG2 (\cite{riek_etal1996}) respectively), in order to make a comparison with HuPrP and MoPrP to further confirm the stability of RaPrP.

In order to get the low pH environment, the residues HIS, ASP, GLU were changed into HIP, ASH, GLH respectively and Cl- ions were added by the XLEaP module of AMBER package. Thus, the salt bridges of the neutral pH environment were broken in low pH environment. 

\section{Results and Discussions}
For seed1$\sim$seed3, at 450 K, during the whole 30 ns, radii of gyrations of the wild-type and the I214V and S173N mutants have been always level off around 15 angstroms whether in neutral pH environment or in low pH environment.

In neutral pH environment, compared with the I214V and S173N mutants (whose three $\alpha$-helices have been unfolded into $\beta$-sheets), the wild-type is always at the lowest level of RMSD (root mean square deviation), RMSF (root mean square fluctuation) and B-factor values. This show to us the wild-type RaPrP is very stable under neutral pH environment. However, under low pH environment (both the wild-type and mutants have been unfolded their structures from $\alpha$-helices into $\beta$-sheets), we cannot see some large differences among the wild-type, the I214V mutant and the S173N mutant.

\subsection{450 K}
The variations of the molecular structures of the wild-type and during the 30 ns for the three sets of MD simulations can be seen in Fig. 1, where the graphs were produced in the use of DSSP program (\cite{kabsch_etal1983}), H is the $\alpha$-helix, I is the $\pi$-helix, G is the 3-helix or 3/10 helix, B is the residue in isolated $\beta$-bridge, E is the extended strand (participates in $\beta$-ladder), T is the hydrogen bonded turn, and S is the bend; H1, H2, and H3 respectively denote the $\alpha$-helix 1 ($\alpha$1), $\alpha$-helix 2 ($\alpha$2) and $\alpha$-helix 3 ($\alpha$3) of a prion protein (we denote the $\beta$-strand before H1 as $\beta$1 and the $\beta$-strand beween H1 and H2 as $\beta$2). We can see that for the wild-type the three $\alpha$-helices have been still kept during the whole 30 ns of each set under the neutral pH environment, but under low pH environment, the three $\alpha$-helices of the wild-type are unfolded as of the I214V and S173N mutants under neutral or low pH environment and at last should be unfolded into rich $\beta$-sheet structures.
\begin{figure}
\centerline{\includegraphics[width=1.2\textwidth]{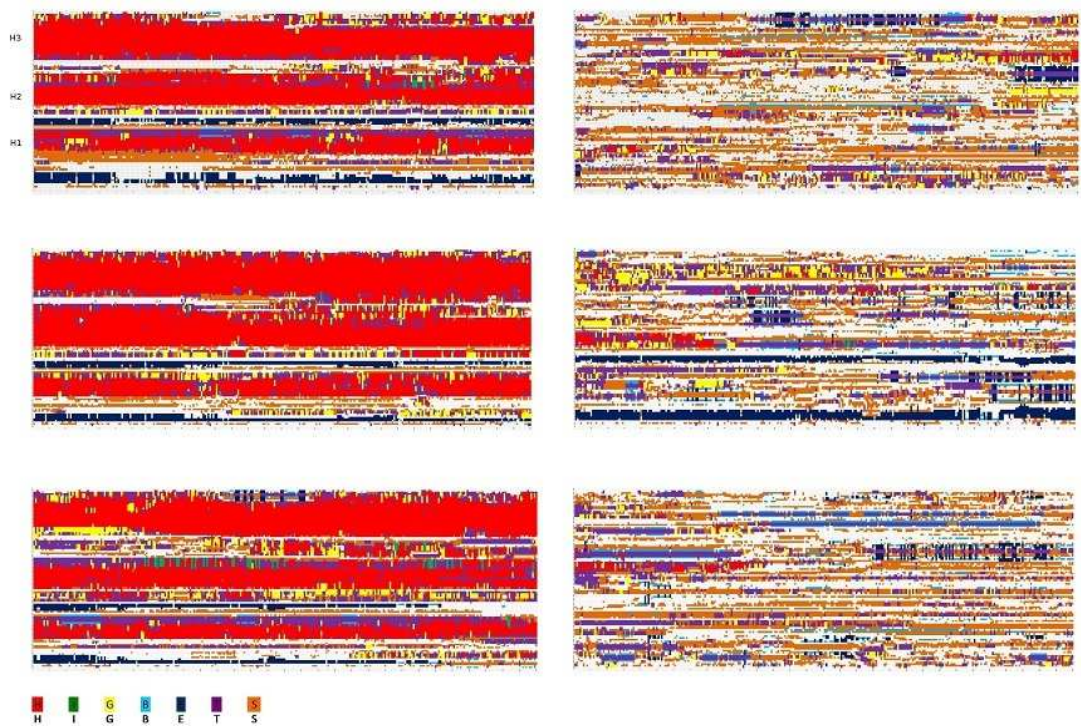}}
\caption{450 K: Secondary Structure graphs for the NMR wild-type RaPrP (seed1 to seed3 from up to down) {\sl [x-axis: time (0--30 ns), y-axis: residue number (124--228); left column: neutral pH, right column: low pH]}.} \label{fig01}
\end{figure}

Under low pH environment, the salt bridges of the wild-type (Table 1) are broken thus leads to the unfolding of the stable helical structures of RaPrP. The following salt bridges should contribute to the stability of RaPrP: {\it {\small ASP146-ARG147, GLU210-ARG207, GLU206-LYS203, GLU206-ARG207, ASP146-HIS139, GLU151-ARG150, GLU151-ARG147, GLU151-ARG155, ASP177-ARG163, GLU145-ARG135, ASP143-HIS139, GLU145-HIS139, GLU195-ARG155, ASP146-ARG150, ASP177-HIS176, ASP143-ARG147, GLU195-LYS193, HIS186-ARG155, HIS186-LYS184, ASP166-ARG163, GLU210-HIS176, HIS139-ARG135, GLU199-LYS193, GLU206-HIS176}}. Among these salt bridges, compared with the salt bridges of I214V and S173N mutants, we find that GLU-195-ARG155 and ASP166-ARG163 are important to contribute to the structural stability of wild-type RaPrP. 
\begin{table}[h]
\centering
{\scriptsize
\begin{tabular}{l             |l                       |l                       |l                       |l} \hline
                Salt bridges  &450 K                   &300 K                   &300 K                   &350 K\\
                Donor-Acceptor&(NMR)                   &(NMR)                   &(X-ray)                 &(NMR)\\ \hline
                ASP146-ARG147 &(100.0, 100.0, 100.0)   &(100.0, 100.0, 100.0)   &(100.0, 100.0, 100.0)   &(100.0, 100.0, 100.0)\\
                GLU210-ARG207 &(99.15, 97.58, 98.43)   &(99.84, 98.01, 47.02)   &(99.95, 99.98, 99.74)   &(99.78, 99.81, 99.84)\\
                GLU206-LYS203 &(97.67, 92.85, 87.35)   &(78.03, 98.85, 91.59)   &(97.67, 99.12, 96.23)   &(82.74, 82.98, 98.35)\\ 
                GLU206-ARG207 &(84.25, 84.80, 75.50)   &(86.90, 65.61, 62.79)   &(81.49, 83.23, 68.71)   &(88.85, 98.84, 89.27)\\
                ASP146-HIS139 &(*, 73.65, *)           &(93.74, 34.67, 44.44)   &(56.70, 90.53, 61.20)   &(92.62, 80.64, 28.38)\\
                GLU151-ARG150 &(72.13, 47.10, 84.25)   &(94.15, 44.07, 29.83)   &(47.79, 26.72, 40.83)   &(76.05, 30.84, 50.65)\\
                GLU151-ARG147 &(63.13, 36.28, 64.22)   &(98.55, 23.69, *)       &(46.27, 42.27, 50.39)   &(*,     31.43, 23.11)\\
                GLU151-ARG155 &(60.78, *, 60.07)       &(56.19, *, 57.29)       &                        &(20.70,     *, 57.71)\\
                ASP177-ARG163 &(47.80, 40.38, 21.92)   &(23.93, 31.62, 82.38)   &(70.59, 26.27, 61.89)   &(19.54,     *, 38.69)\\
                GLU145-ARG135 &(33.52, *, 47.13)       &                        &                        &\\
                ASP143-HIS139 &(27.95, *, 10.95)       &(*, *, 22.43)           &                        &\\
                GLU145-HIS139 &(25.00, *, 18.60)       &                        &                        &\\
                GLU195-ARG155 &(21.88, 7.97, *)        &                        &(13.39, 55.78, 44.30)   &(*,         *, 32.25)\\
                ASP146-ARG150 &(21.23, 66.38, 34.25)   &(92.55, 34.63, 86.09)   &(15.15, *, 10.20)       &(91.38, 62.11, 94.13)\\
                ASP177-HIS176 &(21.17, 16.55, 10.15)   &(15.47, 11.66, *)       &(21.09, 24.35, 19.69)   &(*,     25.63, *)\\
                ASP143-ARG147 &(19.73, 50.30, 28.38)   &(93.43, 42.51, 79.52)   &(53.97, 32.05, 38.03)   &(86.43, 78.38, 73.68)\\
                GLU195-LYS193 &(19.68, 21.98, 10.27)   &(*, *, 25.35)           &(52.20, 61.13, 20.80)   &(*,     15.92, 46.75)\\ 
                HIS186-ARG155 &(14.01, 96.35, 12.47)   &(100.0, 73.69, *)       &(81.19, 35.80, 78.98)   &(71.69, 100.0, *)\\ 
                HIS186-LYS184 &(*, 37.95, 12.95)       &                        &                        &\\ 
                ASP166-ARG163 &(*, *, 54.08)           &                        &                        &\\
                GLU210-HIS176 &(*, *, 33.37)           &                        &                        &(74.31,     *, *)\\ 
                HIS139-ARG135 &(*, *, 32.02)           &                        &                        &\\ 
                GLU199-LYS193 &(*, *, 11.95)           &                        &                        &\\
                GLU199-LYS203 &                        &(11.73, *, *)           &                        &\\
                GLU206-HIS176 &(*, *, 11.83)           &                        &                        &(57.10,     *, 75.83)\\  
                ASP201-ARG155 &(6.62, 6.08, *)         &(3.95, 37.81, *)        &(*, *, 10.05)           &(10.07,     *, *)\\
                GLU199-LYS184 &                        &                        &                        &(46.57,     *, *)\\
                HIS139-ARG150 &                        &                        &                        &(50.96,     *, *)\\
                ASP166-ARG227 &                        &                        &                        &(*,     11.07, *)\\ \hline
\end{tabular}
\caption{Salt bridges of the RaPrP wild-type under neutral pH environment for the MD simulations with occupied rates for seed1, seed2, seed3, where * denotes the occupied rate is less than 10\%.}
}
\end{table}

Zhang (2011) found that there always exist salt bridges between ASP202-ARG156, ASP178-ARG164 in human and mouse prion proteins, between ASP201-ARG155, ASP177-ARG163 in RaPrP (\cite{zhang2011}). The author broke salt bridges of of human, mouse and rabbit prion proteins by doing MD simulations from neutral to low pH environment; consequently, the secondary structures of human and mouse prion proteins were not changed very much, while the stable helical structure of wild-type RaPrP collapsed (\cite{zhang2011}). This is to say that the human and mouse prion protein structures are not affected by removing these salt bridges but the structure of RaPrP is affected very much by these salt bridges. Therefore, compared with human and mouse prion proteins, the salt bridges such as ASP201-ARG155, ASP177-ARG163 contribute greatly to the structural stability of RaPrP.

The salt bridge ASP177-ARG163 is just like a taut bow-string keeping the $\beta$2-$\alpha$2 loop linked. This loop has been a focus on the studies of RaPrP molecular structure (\cite{sweeting_etal2013, sweeting_etal2009, wen_etal2010b, christen_etal2013, christen_etal2012, damberger_etal2011, sigurdson_etal2011, sigurdson_etal2010, pérez_etal2010, christen_etal2009, sigurdson_etal2009, christen_etal2008, gossert_etal2005, lührs_etal2003, stanker_etal2012, cong_etal2013, bett_etal2012, meli_etal2011, rossetti_etal2010, kirby_etal2010, zhang2011_horse, zhang2012}).

Recently, Garrec et al. (2013) reported that the salt linkage of HIS187 and ARG136 of mouse PrP causes ``two misfolding routes for the prion protein" (\cite{garrec_etal2013}). For RaPrP$^\text{C}$, we found that at HIS187 the salt bridge HIS186-ARG155 contributes to the structural stability (Table 1). 

\subsection{At 300 K (Room Temperature)}
We found at the room temperature 300 K, RaPrP has a clear difference from HuPrP and MoPrP: under low pH environment, the three $\alpha$-helices of RaPrP are unfolded for all the seeds (Fig. 2) but for HuPrP and MoPrP the three $\alpha$-helices have not unfolded during the long 30 ns of MD simulations. This implies to us the broken of salt bridges under low pH environment has not affected the structure of HuPrP and MoPrP very much, and indicates to us the C-terminal region of RaPrP$^\text{C}$ has lower thermostability than that of HuPrP$^\text{C}$ and MoPrP$^\text{C}$.
\begin{figure}
\centerline{\includegraphics[width=1.2\textwidth]{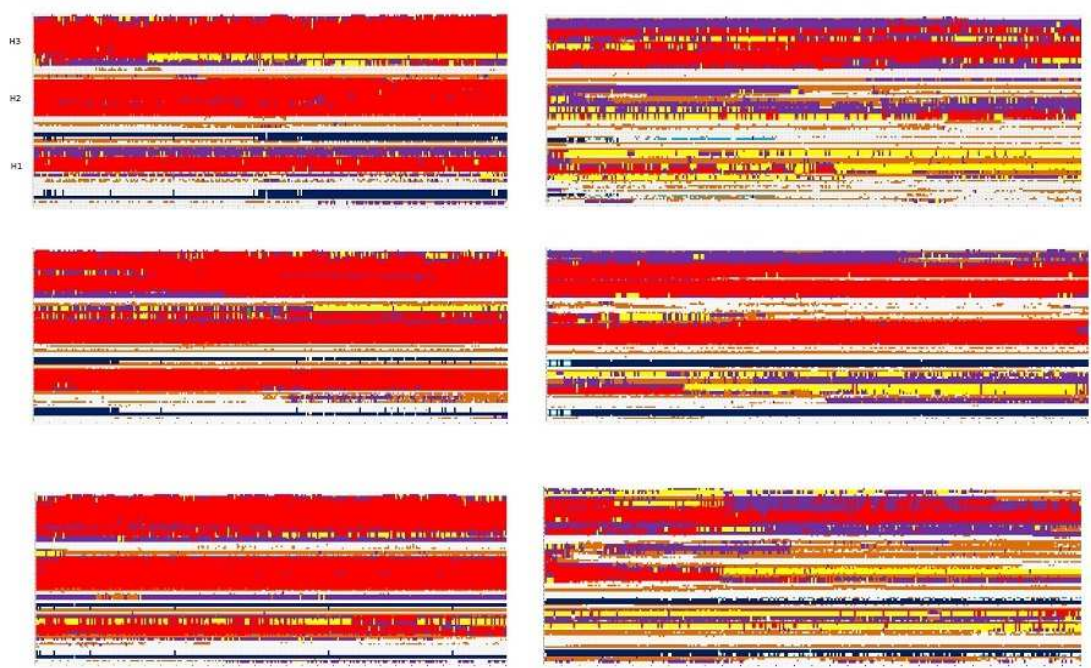}}
\caption{300 K: Secondary Structure graphs for the NMR wild-type RaPrP (seed1 to seed3 from up to down) {\sl [x-axis: time (0--30 ns), y-axis: residue number (124--228); left column: neutral pH, right column: low pH]}.} \label{fig02}
\end{figure}

We also compared with NMR structures with X-ray structures of RaPrP. Generally, for the wild-type RaPrP, under neutral pH environment at 300 K the three $\alpha$-helices have not much unfolded during the whole 30 ns of MD simulations. For the S173N mutant of RaPrP, the secondary structure of X-ray is very similar as that of NMR during the 30 ns of MD simulations at 300 K. The RMSD value between 2FJ3.pdb (NMR, wild-type) and 3O79.pdb (X-ray, wild-type) is 2.791856 angstroms, and between 2JOH.pdb (NMR, S173N mutant) and 4HMM.pdb (X-ray, S173N mutant) is 2.996173 angstroms. Thus, in this paper we used the NMR structures to replace the X-ray structures in our MD result analyses. 

At 300 K, seeing Table 1 we know that the following salt bridges should contribute to the structural stability of RaPrP:\\
$\bullet$ ASP146-ARG147 (in H1),\\ 
$\bullet$ GLU210-ARG207 (in H3),\\
$\bullet$ GLU206-LYS203 (in H3),\\ 
$\bullet$ GLU206-ARG207 (in H3),\\ 
$\bullet$ ASP146-HIS139 (linking H1 $\sim$ loop $\beta$1-$\alpha$1),\\
$\bullet$ GLU151-ARG150 (in H1),\\
$\bullet$ GLU151-ARG147 (in H1),\\ 
$\bullet$ GLU151-ARG155 (in H1),\\ 
$\bullet$ ASP177-ARG163 (linking H2 $\sim$ loop $\beta$2-$\alpha$2),\\
$\bullet$ ASP143-HIS139 (in loop $\beta$1-$\alpha$1),\\
$\bullet$ GLU195-ARG155 (linking loop $\alpha$2-$\alpha$3 $\sim$ bend $\alpha$1-$\beta$2),\\ 
$\bullet$ ASP146-ARG150 (in H1),\\ 
$\bullet$ ASP177-HIS176 (in H2),\\ 
$\bullet$ ASP143-ARG147 (in H1),\\ 
$\bullet$ GLU195-LYS193 (in loop $\alpha$2-$\alpha$3),\\ 
$\bullet$ HIS186-ARG155 (linking C-terminals of H2 and H1),\\
$\bullet$ GLU199-LYS203 (in H3),\\
$\bullet$ ASP201-ARG155 (linking H3 with the C-terminal of H1).\\
All these salt bridge stabilizing interactions added up can make an important contribution to the overall stability of RaPrP$^\text{C}$. Same as at 450 K, we have also found that at 300 K during the long 30 ns the salt bridge ASP177-ARG163 is very strong and always keeps the linking of H2 and the loop $\beta$2-$\alpha$2 (\cite{sweeting_etal2013, sweeting_etal2009, wen_etal2010b, christen_etal2013, christen_etal2012, damberger_etal2011, sigurdson_etal2011, sigurdson_etal2010, pérez_etal2010, christen_etal2009, sigurdson_etal2009, christen_etal2008, gossert_etal2005, lührs_etal2003, stanker_etal2012, cong_etal2013, bett_etal2012, meli_etal2011, rossetti_etal2010, kirby_etal2010, zhang2011_horse, zhang2012}). Observing Fig. 2, we can clearly see that under low pH environment the lose of the salt bridges in H1 made H1 unfolded completely. In Fig. 2, for seed2 under low pH environment, H2 has not unfolded; this almost agrees with the observation from Fig. 3 of (\cite{zhang2011}).

\subsection{At 350 K}
Similarly, we found at temperature 350 K, RaPrP has a clear difference from HuPrP and MoPrP: under low pH environment, the three $\alpha$-helices of RaPrP are unfolded for all the seeds (Fig. 3) but for HuPrP and MoPrP the three $\alpha$-helices have not unfolded during the long 30 ns of MD simulations.
\begin{figure}
\centerline{\includegraphics[width=1.2\textwidth]{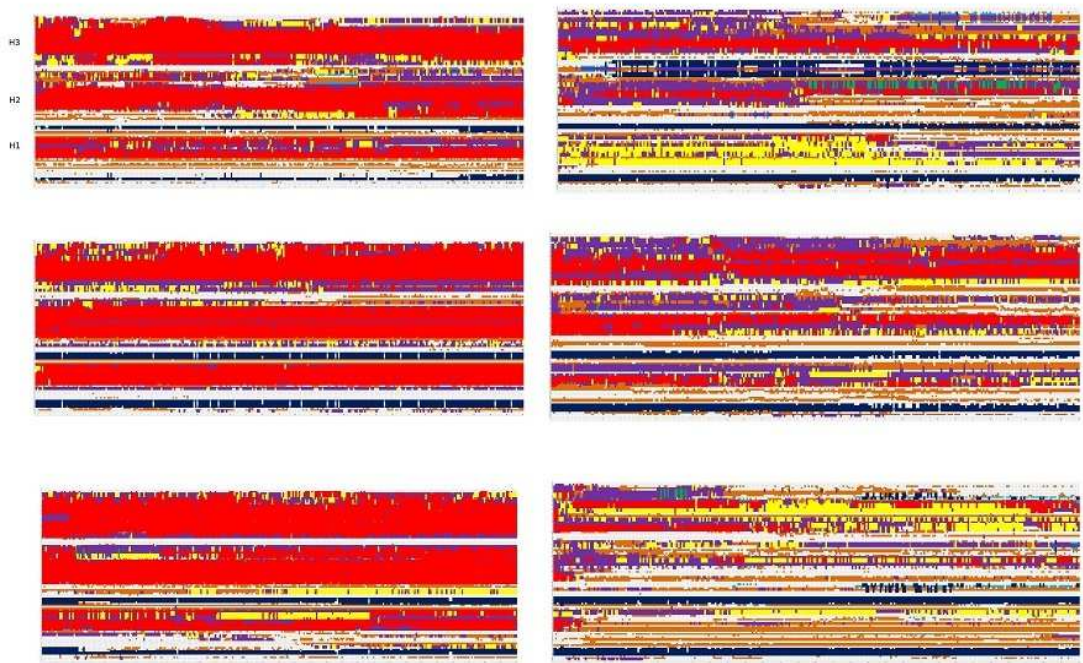}}
\caption{350 K: Secondary Structure graphs for the NMR wild-type RaPrP (seed1 to seed3 from up to down) {\sl [x-axis: time (0--30 ns), y-axis: residue number (124--228); left column: neutral pH, right column: low pH]}.} \label{fig03}
\end{figure}

At 350 K, all the salt bridges listed in Table 1 added up can make an important contribution to the overall stability of RaPrP$^\text{C}$. Same as at 450 K, we have also found that at 350 K during the long 30 ns the salt bridge ASP177-ARG163 is very strong and always keeps the linking of H2 and the loop $\beta$2-$\alpha$2 (\cite{sweeting_etal2013, sweeting_etal2009, wen_etal2010b, christen_etal2013, christen_etal2012, damberger_etal2011, sigurdson_etal2011, sigurdson_etal2010, pérez_etal2010, christen_etal2009, sigurdson_etal2009, christen_etal2008, gossert_etal2005, lührs_etal2003, stanker_etal2012, cong_etal2013, bett_etal2012, meli_etal2011, rossetti_etal2010, kirby_etal2010, zhang2011_horse, zhang2012}). Observing Fig. 3, we can clearly see that under low pH environment the lose of the salt bridges in H1 made H1 unfolded completely, and H2 is almost unfolded. In Fig. 3, for seed3 under low pH environment, H3 is unfolded.

\subsection{Hydrogen Bonds}
A salt bridge is actually a combination of two noncovalent interactions: hydrogen bonding and electrostatic interactions. From neutral pH to low pH, the electrostatic interactions were lost, and it also made the lost of some (weak noncovalent) hydrogen bonding interactions at the same time which made the unfolding of H1$\sim$H3. This is because an $\alpha$-helix is mainly maintained by hydrogen bonding interactions (HBs) and in an ideal $\alpha$-helix there are 3.6 residues per complete rotating so a rotation of 100$^\text{o}$ per residue. We found under neutral pH environment there are HBs ASP177-ARG163, ASP201-ARG155, which have highly occupied during the long 30 ns of MD simulations for seed1$\sim$seed3 at 450 K, 300 K and 350 K (Fig.s 4$\sim$5).
\begin{figure}
\centerline{\includegraphics[width=1.2\textwidth]{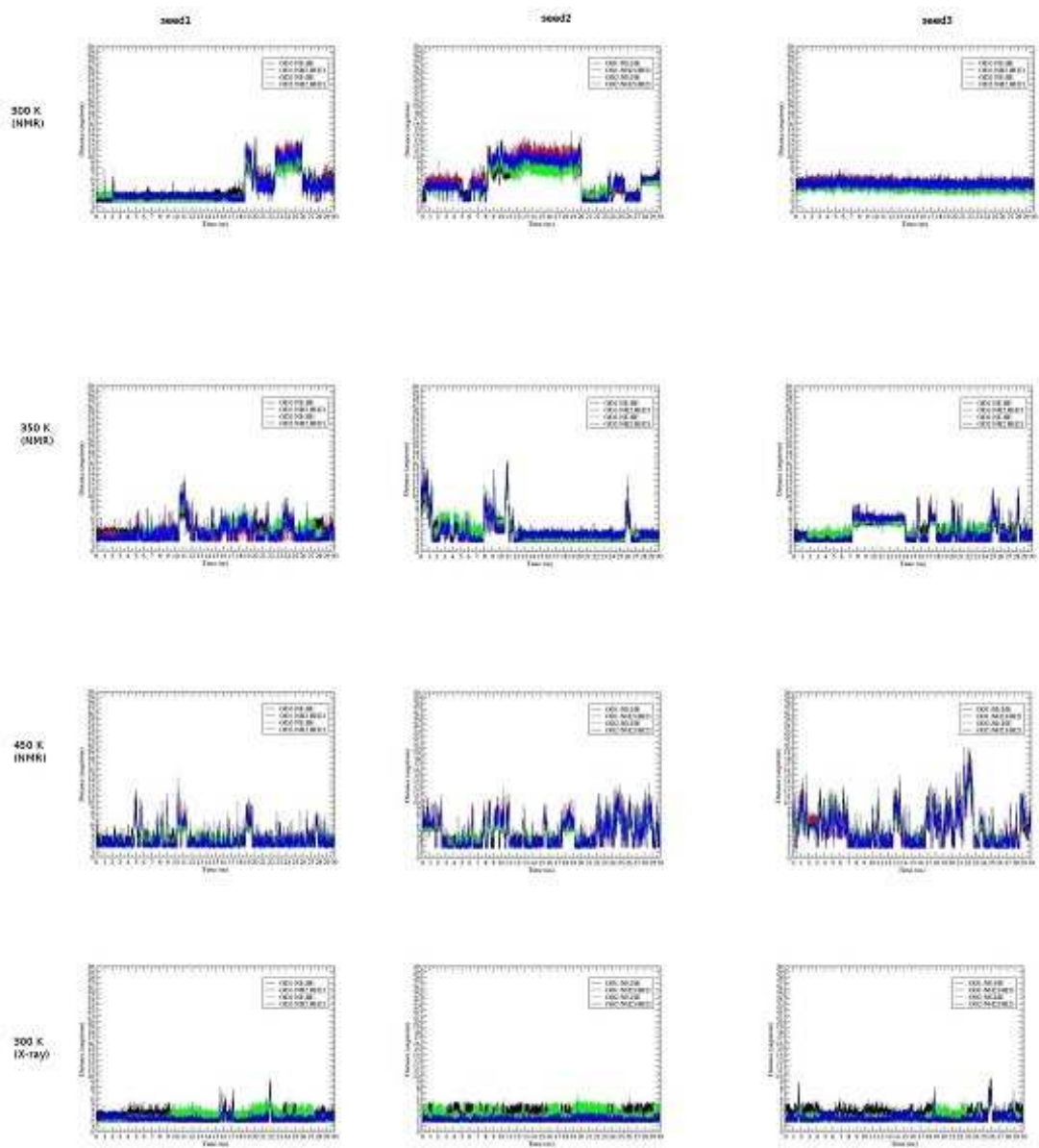}}
\caption{The occupying of HB ASP177-ARG163 during the 30 ns of 300 K (NMR), 350 K (NMR), 450 K (NMR), and 300 K (X-ray) for seed1$\sim$seed3 {\sl [seed1 to seed3: from left to right; 300 K (NMR), 350 K (NMR), 450 K (NMR), and 300 K (X-ray): from up to down]}.} \label{fig04}
\end{figure}
\begin{figure}
\centerline{\includegraphics[width=1.2\textwidth]{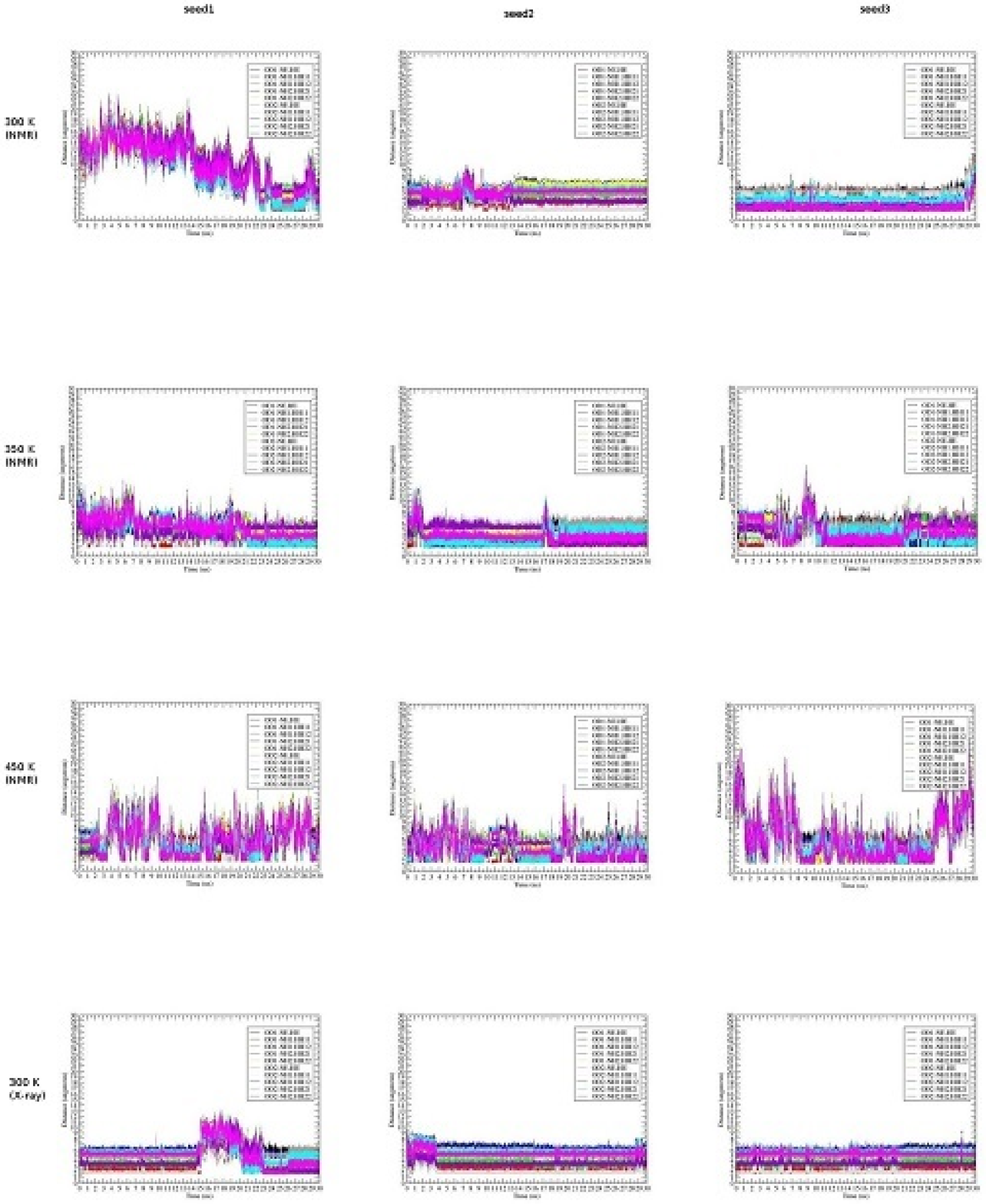}}
\caption{The occupying of HB ASP201-ARG155 during the 30 ns of 300 K (NMR), 350 K (NMR), 450 K (NMR), and 300 K (X-ray) for seed1$\sim$seed3 {\sl [seed1 to seed3: from left to right; 300 K (NMR), 350 K (NMR), 450 K (NMR), and 300 K (X-ray): from up to down]}.} \label{fig05}
\end{figure}

\section{Conclusion}
Like the controversy on ``prion" theory, there is also a big controversy over ``whether rabbits are resistant to prion infection or not?". This paper briefly reviewed the research results on RaPrP and rabbit prions. This paper also did MD simulation studies on RaPrP and its I214V and S173N mutants with some preliminary analyses of the role of salt bridges that plays in the structural stability of RaPrP. The survey shows to us that rabbits were not challenged directly in vivo with other known prion strains and the saPMCA result did not pass the test of the known BSE strain of cattle. Thus, we might still look rabbits as a prion resistant species (and the concept of species resistance to prion disease should be re-evaluated using new powerful tools because it is not reasonable to attribute species-specific prion disease resistance just based purely on the absence of natural cases and incomplete in vivo challenges (\cite{huang_etal2013})). MD results indicate that the three $\alpha$-helices of the wild-type are stable under the neutral pH environment, and the three $\alpha$-helices of the mutants (I214V and S173N) are unfolded into rich $\beta$-sheet structures under the same pH environment. In addition, we found an interesting result that the salt bridges such as ASP201--ARG155, ASP177--ARG163, HIS186-ARG155 contribute greatly to the structural stability of RaPrP.

\section*{Acknowledgments}
{\small This research has been supported by a Victorian Life Sciences Computation Initiative (VLSCI) grant numbered VR0063 on its Peak Computing Facility at the University of Melbourne, an initiative of the Victorian Government of Australia.}

\bibliographystyle{elsarticle-harv}

\begin{thebibliography}{10}
{\small
\bibitem[Barlow and Rennie, 1976]{barlow_etal1976}
Barlow, R.M., Rennie, J.C., 1976. The fate of ME7 scrapie infection in rats, guinea-pigs and rabbits. Res. in Vet. Sci. 21, 110--1.

\bibitem[Baron et al., 1988]{baron_etal1988}
Baron, H., Baron-Van Evercooren, A., Brucher, J.M., 1988. Antiserum to scrapie-associated fibril protein reacts with amyloid plaques in familial transmissible dementia. J. Neuropathol. Exp. Neurol. 47(2), 158--65.

\bibitem[Barry et al., 1986]{barry_etal1986}
Barry, R.A., Kent, S.B., McKinley, M.P., Meyer, R.K., DeArmond, S.J., Hood, L.E., Prusiner, S.B., 1986. Scrapie and cellular prion proteins share polypeptide epitopes. J. Infect. Dis. 153(5), 848--54.

\bibitem[Barry et al., 1985]{barry_etal1985}
Barry, R.A., McKinley, M.P., Bendheim, P.E., Lewis, G.K., DeArmond, S.J., Prusiner, S.B., 1985. Antibodies to the scrapie protein decorate prion rods. J. Immunol. 135(1), 603--13.

\bibitem[Barry et al., 1988]{barry_etal1988}
Barry, R.A., Vincent, M.T., Kent, S.B., Hood, L.E., Prusiner, S.B., 1988. Characterization of prion proteins with monospecific antisera to synthetic peptides. J. Immunol. 140(4), 1188--93.

\bibitem[Bastian et al., 2007]{bastian_etal2007}
Bastian, F.O., Sanders, D.E., Forbes, W.A., Hagius, S.D., Walker, J.V., Henk, W.G., Enright, F.M., Elzer, P.H., 2007. Spiroplasma spp. from transmissible spongiform encephalopathy brains or ticks induce spongiform encephalopathy in ruminants. J. Med. Microbiol. 56(Pt 9), 1235--42.

\bibitem[Bencsik et al., 2001]{bencsik_etal2001}
Bencsik, A., Lezmi, S., Hunsmann, G., Baron, T., 2001. Close vicinity of PrP expressing cells (FDC) with noradrenergic fibers in healthy sheep spleen. Dev. Immunol. 8(3--4), 235--41.

\bibitem[Bendheim et al., 1984]{bendheim_etal1984}
Bendheim, P.E., Barry, R.A., DeArmond, S.J., Stites, D.P., Prusiner, S.B., 1984. Antibodies to a scrapie prion protein. Nature 310(5976), 418--21.

\bibitem[Bett et al., 2012]{bett_etal2012}
Bett, C., Fernández-Borges, N., Kurt, T.D., Lucero, M., Nilsson, K.P., Castilla, J., Sigurdson, C.J., 2012. Structure of the $\beta$2-$\alpha$2 loop and interspecies prion transmission. FASEB. J. 26(7), 2868--76.

\bibitem[Biswas et al., 2006]{biswas_etal2006}
Biswas, S., Langeveld, J.P., Tipper, D., Lu, S., 2006. Intracellular accumulation of a 46 kDa species of mouse prion protein as a result of loss of glycosylation in cultured mammalian cells. Biochem. Biophys. Res. Commun. 349(1), 153--61.

\bibitem[Bitel et al., 2010]{bitel_etal2010}
Bitel, C.L., Feng, Y., Souayah, N., Frederikse, P.H., 2010. Increased expression and local accumulation of the prion protein, Alzheimer Aß peptides, superoxide dismutase 1, and nitric oxide synthases 1 \& 2 in muscle in a rabbit model of diabetes. BMC Physiol 10, 18. doi: 10.1186/1472-6793-10-18.

\bibitem[Bjorndahl et al., 2011]{bjorndahl_etal2011}
Bjorndahl, T.C., Zhou, G.P., Liu, X.H., Pineiro, R.P., Semenchenko, V., Saleem, F., Acharya, S., Bujold, A., Sobsey, C.A., Wishart,  D.S., 2011. Detailed  biophysical  characterization  of  the acid-induced  PrP$^\text{C}$   to  PrP  conversion  process. Biochemistry 50(7), 1162–-73.

\bibitem[Bockman et al., 1987]{bockman_etal1987}
Bockman, J.M., Prusiner, S.B., Tateishi, J., Kingsbury, D.T., 1987. Immunoblotting of Creutzfeldt-Jakob disease prion proteins: host species-specific epitopes. Ann. Neurol. 21(6), 589--95.

\bibitem[Bode et al., 1985]{bode_etal1985}
Bode, L., Pocchiari, M., Gelderblom, H., Diringer, H., 1985. Characterization of antisera against scrapie-associated fibrils (SAF) from affected hamster and cross-reactivity with SAF from scrapie-affected mice and from patients with Creutzfeldt-Jakob disease. J. Gen. Virol. 66(Pt 11), 2471--8.

\bibitem[Brun et al., 2004]{brun_etal2004}
Brun, A., Castilla, J., Ramírez, M.A., Prager, K., Parra, B., Salguero, F.J., Shiveral, D., Sánchez, C., Sánchez-Vizcaíno, J.M., Douglas, A., Torres, J.M., 2004. Proteinase K enhanced immunoreactivity of the prion protein-specific monoclonal antibody 2A11. Neurosci. Res. 48(1), 75--83.

\bibitem[Caughey et al., 1988]{caughey_etal1988}
Caughey, B., Race, R., Vogel, M., Buchmeier, M., Chesebro, B., 1988. In vitro expression of cloned PrP cDNA derived from scrapie-infected mouse brain: lack of transmission of scrapie infectivity. Ciba. Found. Symp. 135, 197--208.

\bibitem[Chianini et al., 2012]{chianini_etal2012}
Chianini, F., Fernndez-Borges, N., Vidal, E., Gibbard, L., Pintado, B., de Castro, J., Priola, S.A., Hamilton, S., Eatona, L.S., Finlayson, J., Pang, Y., Steele, P., Reid, H.W., Dagleish, M.P., Castilla, J., 2012. Rabbits are not resistant to prion infection. Proc. Natl. Acad. Sci.  U. S. A. 109(13) : 5080--5.

\bibitem[Cho, 1986]{cho1986}
Cho HJ, 1986. Antibody to scrapie-associated fibril protein identifies a cellular antigen. J. Gen. Virol. 67(Pt 2): 243--53.

\bibitem[Chou, 1987]{chou1987}
Chou, K.C., 1987. The biological functions of low-frequency phonons: 6. A possible dynamic mechanism of allosteric transition in antibody molecules. Biopolymers 26, 285--95.

\bibitem[Chou, 1988]{chou1988}
Chou, K.C., 1988. Review: Low-frequency collective motion in biomacromolecules and its biological functions. Biophys. Chem. 30, 3--48.

\bibitem[Chou, 1989]{chou1989}
Chou, K.C., 1989. Low-frequency resonance and cooperativity of hemoglobin. Trends Biochem. Sci. 14, 212--3.

\bibitem[Chou et al., 2011]{chou_etal2011}
Chou, K.C., Lin, W.Z.,  Xiao, X., 2011.  Wenxiang:  a  web-server  for drawing wenxiang diagrams. Nat. Sci. 3, 862--5.  

\bibitem[Chou and Mao, 1988]{choum1988}
Chou, K.C., and Mao, B., 1988. Collective motion in DNA and its role in drug intercalation. Biopolymers 27, 1795--815.

\bibitem[Chou et al., 1994]{chou_etal1994}
Chou, K.C., Zhang, C.T., Maggiora, G.M., 1994. Solitary wave dynamics as a mechanism for explaining the internal motion during microtubule growth. Biopolymers 34, 143--53.

\bibitem[Chou et al., 1997]{chou_etal1997}
Chou, K.C., Zhang, C.T., Maggiora, G.M., 1997. Disposition of amphiphilic helices  in  heteropolar  environments.  Proteins:  Struct. Funct. Genet. 28, 99--108. 

\bibitem[Christen et al., 2013]{christen_etal2013}
Christen, B., Damberger, F.F., Pérez, D.R., Hornemann, S., Wüthrich, K., 2013. Structural plasticity of the cellular prion protein and implications in health and disease. Proc. Natl. Acad. Sci. U. S. A. 110(21), 8549--54.

\bibitem[Christen et al., 2009]{christen_etal2009}
Christen, B., Hornemann, S., Damberger, F.F., Wüthrich, K., 2009. Prion protein NMR structure from tammar wallaby (Macropus eugenii) shows that the $\beta$2-$\alpha$2 loop is modulated by long-range sequence effects. J. Mol. Biol. 389(5), 833--45.

\bibitem[Christen et al., 2012]{christen_etal2012}
Christen, B., Hornemann, S., Damberger, F.F., Wüthrich, K., 2012. Prion protein mPrP [F175A](121--231): structure and stability in solution. J. Mol. Biol. 423(4), 496--502.

\bibitem[Christen et al., 2008]{christen_etal2008}
Christen, B., Pérez, D.R., Hornemann, S., Wüthrich, K., 2008. NMR structure of the bank vole prion protein at 20 degrees C contains a structured loop of residues 165--171. J. Mol. Biol. 383(2), 306--12.

\bibitem[Cong et al., 2013]{cong_etal2013}
Cong, X., Bongarzone, S., Giachin, G., Rossetti, G., Carloni, P., Legname, G., 2013. Dominantnegative effects in prion diseases: insights from molecular dynamics simulations on mouse prion protein chimeras. J. Biomol. Struct. Dyn. 31(8), 829--40.

\bibitem[Courageot et al., 2008]{courageot_etal2008}
Courageot, M.P., Daude, N., Nonno, R., Paquet, S., Di Bari, M.A., Le Dur, A., Chapuis, J., Hill, A.F., Agrimi, U., Laude, H., Vilette, D., 2008. A cell line infectible by prion strains from different species. J. Gen. Virol. 89(Pt 1), 341--7.

\bibitem[Damberger et al., 2011]{damberger_etal2011}
Damberger, F.F., Christen, B., Pérez, D.R., Hornemann, S., Wüthrich, K., 2011. Cellular prion protein conformation and function. Proc. Natl. Acad. Sci. U. S. A. 108(42), 17308--13.

\bibitem[Di Martino et al., 1991]{dimartino_etal1991}
Di Martino, A., Bigon, E., Corona, G., Callegaro, L., 1991. Production and characterization of antibodies to mouse scrapie-amyloid protein elicited by non-carrier linked synthetic peptide immunogens. J. Mol. Recognit. 4(2--3), 85--91.

\bibitem[Dong et al., 2007]{dong_etal2007}
Dong, C.F., Wang, X.F., An, R., Chen, J.M., Shan, B., Han, L., Lei, Y.J., Han, J., Dong, X.P., 2007. [Interaction analysis between various PrP fusion proteins and the tubulin in vitro]. Bing Du Xue Bao 23(1), 28--32.

\bibitem[Dupiereux et al., 2006]{dupiereux_etal2006}
Dupiereux, I., Zorzi, W., Rachidi, W., Zorzi, D., Pierard, O., Lhereux, B., Heinen, E., Elmoualij, B., 2006. Study on the toxic mechanism of prion protein peptide 106-126 in neuronal and non neuronal cells. J. Neurosci. Res. 84(3), 637--46.

\bibitem[Farquhar et al., 1989]{farquhar_etal1989}
Farquhar, C.F., Somerville, R.A., Ritchie, L.A., 1989. Post-mortem immunodiagnosis of scrapie and bovine spongiform encephalopathy. J. Virol. Methods 24(1-2), 215--21.

\bibitem[Fernández-Borges et al., 2012]{fernandez-borges_etal2012}
Fernández-Borges, N., Chianini, F., Eraña, H., Vidal, E., Eaton, S.L., Pintado, B., Finlayson, J., Dagleish, M.P., Castilla, J., 2012. Naturally prion resistant mammals: a utopia? Prion 6(5), 425--9.

\bibitem[Fernandez-Funez et al., 2009]{fernandez-funez_etal2009}
Fernandez-Funez, P., Casas-Tinto, S., Zhang, Y., Gomez-Velazquez, M., Morales-Garza, M.A., Cepeda-Nieto, A.C., Castilla, J., Soto, C., Rincon-Limas, D.E., 2009. In vivo generation of neurotoxic prion protein: role for hsp70 in accumulation of misfolded isoforms. PLoS Genet 5(6), e1000507.

\bibitem[Fernandez-Funez et al., 2010]{fernandez-funez_etal2010}
Fernandez-Funez, P., Zhang, Y., Casas-Tinto, S., Xiao, X., Zou, W.Q., Rincon-Limas, D.E., 2010. Sequence-dependent prion protein misfolding and neurotoxicity. J. Biol. Chem. 285(47), 36897--908.

\bibitem[Fernandez-Funez et al., 2011]{fernandez-funez_etal2011}
Fernandez-Funez, P., Zhang, Y., Sanchez-Garcia, J., Jensen, K., Zou, W.Q., Rincon-Limas, D.E., 2011. Pulling rabbits to reveal the secrets of the prion protein. Commun. Integr. Biol. 4(3), 262--6.

\bibitem[Friedman-Levi et al., 2013]{friedman-levi_etal2013}
Friedman-Levi, Y., Mizrahi, M., Frid, K., Binyamin, O., Gabizon, R., 2013. PrPST, a Soluble, Protease Resistant and Truncated PrP Form Features in the Pathogenesis of a Genetic Prion Disease. PLoS ONE 8(7), e69583.

\bibitem[Gabizon et al., 1988]{gabizon_etal1988}
Gabizon, R., McKinley, M.P., Groth, D., Prusiner, S.B., 1988. Immunoaffinity purification and neutralization of scrapie prion infectivity. Proc. Natl. Acad. Sci. U. S. A. 85(18), 6617--21.

\bibitem[Gabizon et al., 1989]{gabizon_etal1989}
Gabizon, R., McKinley, M.P., Groth, D., Westaway, D., DeArmond, S.J., Carlson, G.A., Prusiner, S.B., 1989. Immunoaffinity purification and neutralization of scrapie prions. Prog. Clin. Biol. Res. 317, 583--600.

\bibitem[Gao et al., 2006]{gao_etal2006}
Gao, C., Lei, Y.J., Han, J., Shi, Q., Chen, L., Guo, Y., Gao, Y.J., Chen, J.M., Jiang, H.Y., Zhou, W., Dong, X.P., 2006. Recombinant neural protein PrP can bind with both recombinant and native apolipoprotein E in vitro. Acta. Biochim. Biophys. Sin. (Shanghai) 38(9), 593--601.

\bibitem[Garssen et al., 2000]{garssen_etal2000}
Garssen, G.J., Van Keulen, L.J., Farquhar, C.F., Smits, M.A., Jacobs, J.G., Bossers, A., Meloen, R.H., Langeveld, J.P., 2000. Applicability of three anti-PrP peptide sera including staining of tonsils and brainstem of sheep with scrapie. Microsc. Res. Tech. 50(1), 32--9.

\bibitem[Gerber et al., 2008]{gerber_etal2008}
Gerber, R.,  Tahiri-Alaoui, A.,  Hore, P.J.,  James, W., 2008. Conformational pH dependence of intermediate states during oligomerization of the human prion protein. Protein Sci. 17(3), 537--44.

\bibitem[Gilch et al., 2003]{gilch_etal2003}
Gilch, S., Wopfner, F., Renner-Müller, I., Kremmer, E., Bauer, C., Wolf, E., Brem, G., Groschup, M.H., Schätzl, H.M., 2003. Polyclonal anti-PrP auto-antibodies induced with dimeric PrP interfere efficiently with PrPSc propagation in prion-infected cells. J. Biol. Chem. 278(20), 18524--31.

\bibitem[Golanska et al., 2005]{golanska_etal2005}
Golanska, E., Hulas-Bigoszewska, K., Sikorska, B., Liberski, P.P., 2005. [Analyses of 14--3--3 protein in the cerebrospinal fluid in Creutzfeldt-Jakob disease. Preliminary report]. Neurol. Neurochir. Pol. 39(5), 358--65.

\bibitem[Gossert et al., 2005]{gossert_etal2005}
Gossert, A.D., Bonjour, S., Lysek, D.A., Fiorito, F., Wüthrich, K., 2005. Prion protein NMR structures of elk and of mouse/elk hybrids. Proc. Natl. Acad. Sci. U. S. A. 102(3), 646--50.

\bibitem[Garrec et al., 2013]{garrec_etal2013}
Garrec, J., Tavernelli, I., Rothlisberger, U., 2013. Two misfolding routes for the prion protein around pH 4.5. PLoS Comput. Biol. 9(5), e1003057.

\bibitem[Groschup et al., 1997]{groschup_etal1997}
Groschup, M.H., Harmeyer, S., Pfaff, E., 1997. Antigenic features of prion proteins of sheep and of other mammalian species. J. Immunol. Methods 207(1), 89--101.

\bibitem[Groschup et al., 1994]{groschup_etal1994}
Groschup, M.H., Langeveld, J., Pfaff, E., 1994. The major species specific epitope in prion proteins of ruminants. Arch. Virol. 136(3-4), 423--31.

\bibitem[Groschup and Pfaff, 1993]{groschupp1993}
Groschup, M.H., Pfaff, E., 1993. Studies on a species-specific epitope in murine, ovine and bovine prion protein. J. Gen. Virol. 74(Pt 7), 1451--6.

\bibitem[Guiroy et al., 1993]{guiroy_etal1993}
Guiroy, D.C., Marsh, R.F., Yanagihara, R., Gajdusek, D.C., 1993. Immunolocalization of scrapie amyloid in non-congophilic, non-birefringent deposits in golden Syrian hamsters with experimental transmissible mink encephalopathy. Neurosci. Lett. 155(1), 112--5.

\bibitem[Handisurya et al., 2007]{handisurya_etal2007}
Handisurya, A., Gilch, S., Winter, D., Shafti-Keramat, S., Maurer, D., Schätzl, H.M., Kirnbauer, R., 2007. Vaccination with prion peptide-displaying papillomavirus-like particles induces autoantibodies to normal prion protein that interfere with pathologic prion protein production in infected cells. FEBS. J. 274(7), 1747--58.

\bibitem[Hanoux et al., 2009]{hanoux_etal2009}
Hanoux, V., Wijkhuisen, A., Alexandrenne, C., Créminon, C., Boquet, D., Couraud, J.Y., 2009. Polyclonal anti-idiotypic antibodies which mimic an epitope of the human prion protein. Mol. Immunol. 46(6), 1076--83.

\bibitem[Hashimoto et al., 1992]{hashimoto_etal1992}
Hashimoto, K., Mannen, T., Nukina, N., 1992. Immunohistochemical study of kuru plaques using antibodies against synthetic prion protein peptides. Acta. Neuropathol. 83(6), 613--7.

\bibitem[Hay et al., 1987]{hay_etal1987}
Hay, B., Prusiner, S.B., Lingappa, V.R., 1987. Evidence for a secretory form of the cellular prion protein. Biochemistry 26(25), 8110--5.

\bibitem[Huang et al., 2013]{huang_etal2013}
Huang, P., Lian, F., Wen, Y., Guo, C., Lin D., 2013. Prion protein oligomer and its neurotoxicity. Acta. Biochim. Biophys. Sin. (Shanghai) 45(6), 442--51. 

\bibitem[Ikegami et al., 1991]{ikegami_etal1991}
Ikegami, Y., Ito, M., Isomura, H., Momotani, E., Sasaki, K., Muramatsu, Y., Ishiguro, N., Shinagawa, M., 1991. Pre-clinical and clinical diagnosis of scrapie by detection of PrP protein in tissues of sheep. Vet. Rec. 128(12), 271--5.

\bibitem[Ishiguro et al., 2009]{ishiguro_etal2009}
Ishiguro, N., Inoshima, Y., Sassa, Y., Takahashi, T., 2009. Molecular characterization of chicken prion proteins by C-terminal-specific monoclonal antibodies. Vet. Immunol. Immunopathol. 128(4), 402--6.

\bibitem[Ishikawa and Kuwata, 2010]{ishikawa_etal2010}
Ishikawa, T., Kuwata, K., 2010. Interaction analysis of the native structure of  prion  protein  with  quantum  chemical  calculations.  J.  Chem. Theor. Comput. 6, 538--47. 

\bibitem[Jackman and Schmerr, 2003]{jackmans2003}
Jackman, R., Schmerr, M.J., 2003. Analysis of the performance of antibody capture methods using fluorescent peptides with capillary zone electrophoresis with laser-induced fluorescence. Electrophoresis 24(5), 892--6.

\bibitem[Julien et al., 2011]{julien_etal2011}
Julien, O., Chatterjee, S., Bjorndahl, T.C., Sweeting, B., Acharya, S., Semenchenko, V., Chakrabartty, A., Pai, E.F., Wishart, D.S., Sykes, B.D., Cashman, N.R., 2011. Relative and regional stabilities of the hamster, mouse, rabbit, and bovine prion proteins toward urea unfolding assessed by nuclear magnetic resonance and circular dichroism spectroscopies. Biochemistry 50(35). 7536--45.

\bibitem[Kabsch and Sander, 1983]{kabsch_etal1983}
Kabsch, W., and Sander, C., 1983. Dictionary of protein secondary structure: pattern recognition of hydrogen-bonded and geometrical features. Biopolymers 22, 2577–-637.

\bibitem[Kascsak et al., 1986]{kascsak_etal1986}
Kascsak, R.J., Rubenstein, R., Merz, P.A., Carp, R.I., Robakis, N.K., Wisniewski, H.M., Diringer, H., 1986. Immunological comparison of scrapie-associated fibrils isolated from animals infected with four different scrapie strains. J Virol 59(3), 676--83.

\bibitem[Kascsak et al., 1987]{kascsak_etal1987}
Kascsak, R.J., Rubenstein, R., Merz, P.A., Tonna-DeMasi, M., Fersko, R., Carp, R.I., Wisniewski, H.M., Diringer, H., 1987. Mouse polyclonal and monoclonal antibody to scrapie-associated fibril proteins. J. Virol. 61(12), 3688--93.

\bibitem[Kelker et al., 2001]{kelker_etal2001}
Kelker, M., Kim, C., Chueh, P.J., Guimont, R., Morré, D.M., Morré, D.J., 2001. Cancer isoform of a tumor-associated cell surface NADH oxidase (tNOX) has properties of a prion. Biochemistry 40(25), 7351--4.

\bibitem[Khan et al., 2010]{khan_etal2010}
Khan, M.Q., Sweeting, B., Mulligan, V.K., Arslan, P.E., Cashman, N.R., Pai, E.F., Chakrabartty, A., 2010. Prion disease susceptibility is affected by $\beta$-structure folding propensity and local side-chain interactions in PrP.  Proc. Natl. Acad. Sci. U. S. A. 107(46), 19808--13.

\bibitem[Kim et al., 2012]{kim_etal2012}
Kim, H.J., Tark, D.S., Lee, Y.H., Kim, M.J., Lee, W.Y., Cho, I.S., Sohn, H.J., Yokoyama, T., 2012. Establishment of a cell line persistently infected with chronic wasting disease prions. J. Vet. Med. Sci. 74(10), 1377--80.

\bibitem[Kirby et al., 2010]{kirby_etal2010}
Kirby, L., Agarwal, S., Graham, J.F., Goldmann, W., Gill, A.C., 2010. Inverse correlation of thermal lability and conversion efficiency for five prion protein polymorphic variants. Biochemistry 49(7), 1448--59.

\bibitem[Kirkwood et al., 1992]{kirkwood_etal1992}
Kirkwood, J.K., Wells, G.A., Cunningham, A.A., Jackson, S.I., Scott, A.C., Dawson, M., Wilesmith, J.W., 1992. Scrapie-like encephalopathy in a greater kudu (Tragelaphus strepsiceros) which had not been fed ruminant-derived protein. Vet. Rec. 130(17), 365--7.

\bibitem[Kitamoto et al., 1989]{kitamoto_etal1989}
Kitamoto, T., Tateishi, J., Sawa, H., Doh-Ura, K., 1989. Positive transmission of Creutzfeldt-Jakob disease verified by murine kuru plaques. Lab. Invest. 60(4), 507--12.

\bibitem[Kocisko and Caughey, 2006]{kociskoc2006}
Kocisko, D.A., Caughey, B., 2006. Searching for anti-prion compounds: cell-based high-throughput in vitro assays and animal testing strategies. Methods Enzymol. 412, 223--34.

\bibitem[Komar et al., 1997]{komar_etal1997}
Komar, A.A., Lesnik, T., Cullin, C., Guillemet, E., Ehrlich, R., Reiss, C., 1997. Differential resistance to proteinase K digestion of the yeast prion-like (Ure2p) protein synthesized in vitro in wheat germ extract and rabbit reticulocyte lysate cell-free translation systems. FEBS. Lett. 415(1), 6--10.

\bibitem[Korth et al., 1997]{korth_etal1997}
Korth, C., Stierli, B., Streit, P., Moser, M., Schaller, O., Fischer, R., Schulz-Schaeffer, W., Kretzschmar, H., Raeber, A., Braun, U., Ehrensperger, F., Hornemann, S., Glockshuber, R., Riek, R., Billeter, M., Wthrich, K., Oesch, B., 1997. Prion (PrPSc)-specific epitope defined by a monoclonal antibody. Nature 390(6655), 74--7.

\bibitem[Laude et al., 2002]{laude_etal2002}
Laude, H., Vilette, D., Le Dur, A., Archer, F., Soulier, S., Besnard, N., Essalmani, R., Vilotte, J.L., 2002. New in vivo and ex vivo models for the experimental study of sheep scrapie: development and perspectives. C. R. Biol. 325(1), 49--57.

\bibitem[Lawson et al., 2008]{lawson_etal2008}
Lawson, V.A., Vella, L.J., Stewart, J.D., Sharples, R.A., Klemm, H., Machalek, D.M., Masters, C.L., Cappai, R., Collins, S.J., Hill, A.F., 2008. Mouse-adapted sporadic human Creutzfeldt-Jakob disease prions propagate in cell culture. Int. J. Biochem. Cell. Biol. 40(12), 2793--801.

\bibitem[Li et al., 2007]{li_etal2007}
Li, J., Mei, F.H., Xiao, G.F., Guo, C.Y., Lin, D.H. 2007. 1$^{\text{H}}$, 13$^{\text{C}}$ and 15$^{\text{N}}$ resonance assignments of rabbit prion protein 91--228. J. Biomol. NMR 38, pp. 181.

\bibitem[Li et al., 2001]{li_etal2001}
Li, Y.M., Tian, B., Zhang, B.Y., Dong, X.P., 2001. [Diagnosis of bovine spongiform encephalopathy and scrapie by Western blot]. Sheng Wu Gong Cheng Xue Bao 17(5), 494--7.

\bibitem[Loftus and Rogers, 1997]{loftusr1997}
Loftus, B., Rogers, M., 1997. Characterization of a prion protein (PrP) gene from rabbit; a species with apparent resistance to infection by prions. Gene 184(2), 215--9.

\bibitem[Lopez et al., 1990]{lopez_etal1990}
Lopez, C.D., Yost, C.S., Prusiner, S.B., Myers, R.M., Lingappa, V.R., 1990. Unusual topogenic sequence directs prion protein biogenesis. Science 248(4952), 226--9.

\bibitem[Lührs et al., 2003]{lührs_etal2003}
Lührs, T., Riek, R., Güntert, P., Wüthrich, K., 2003. NMR structure of the human doppel protein. J. Mol. Biol. 326(5), 1549--57.

\bibitem[Ma et al., 2012]{ma_etal2012}
Ma, Q., Fan, J.B., Zhou, Z., Zhou, B.R., Meng, S.R., Hu, J.Y., Chen, J., Liang, Y., 2012. The contrasting effect of macromolecular crowding on amyloid fibril formation. PLoS One 7(4), e36288.

\bibitem[Madec et al., 1997]{madec_etal1997}
Madec, J.Y., Vanier, A., Dorier, A., Bernillon, J., Belli, P., Baron, T., 1997. Biochemical properties of protease resistant prion protein PrPSc in natural sheep scrapie. Arch. Virol. 142(8), 1603--12.

\bibitem[Mays et al., 2011]{mays_etal2011}
Mays, C.E., Yeom, J., Kang, H.E., Bian, J., Khaychuk, V., Kim, Y., Bartz, J.C., Telling, G.C., Ryou, C., 2011. In vitro amplification of misfolded prion protein using lysate of cultured cells. PLoS One 6(3), e18047.

\bibitem[Meli et al., 2011]{meli_etal2011}
Meli, M., Gasset, M., Colombo, G., 2011. Dynamic diagnosis of familial prion diseases supports the $\beta$2-$\alpha$2 loop as a universal interference target. PLoS One 6(4), e19093.

\bibitem[Merz et al., 1987]{merz_etal1987}
Merz, P.A., Kascsak, R.J., Rubenstein, R., Carp, R.I., Wisniewski, H.M., 1987. Antisera to scrapie-associated fibril protein and prion protein decorate scrapie-associated fibrils. J. Virol. 61(1), 42--9.

\bibitem[Miller et al., 1993]{miller_etal1993}
Miller, J.M., Jenny, A.L., Taylor, W.D., Marsh, R.F., Rubenstein, R., Race, R.E., 1993. Immunohistochemical detection of prion protein in sheep with scrapie. J. Vet. Diagn. Invest. 5(3), 309--16.

\bibitem[Nisbet et al., 2010]{nisbet_etal2010}
Nisbet, R.M., Harrison, C.F., Lawson, V.A., Masters, C.L., Cappai, R., Hill, A.F., 2010. Residues surrounding the glycosylphosphatidylinositol anchor attachment site of PrP modulate prion infection: insight from the resistance of rabbits to prion disease. J. Virol. 84(13), 6678--86.

\bibitem[Oboznaya et al., 2007]{oboznaya_etal2007}
Oboznaya, M.B., Gilch, S., Titova, M.A., Koroev, D.O., Volkova, T.D., Volpina, O.M., Schätzl, H.M., 2007. Antibodies to a nonconjugated prion protein peptide 95--123 interfere with PrPSc propagation in prion-infected cells. Cell. Mol. Neurobiol. 27(3), 271--84.

\bibitem[Onodera et al., 1993]{onodera_etal1993}
Onodera, T., Ikeda, T., Muramatsu, Y., Shinagawa, M., 1993. Isolation of scrapie agent from the placenta of sheep with natural scrapie in Japan. Microbiol. Immunol. 37(4), 311--6.

\bibitem[Pérez et al., 2010]{pérez_etal2010}
Pérez, D.R., Damberger, F.F., Wüthrich, K., 2010. Horse prion protein NMR structure and comparisons with related variants of the mouse prion protein. J. Mol. Biol. 400(2), 121--8.

\bibitem[Polymenidou et al., 2008]{polymenidou_etal2008}
Polymenidou, M., Trusheim, H., Stallmach, L., Moos, R., Julius, C., Miele, G., Lenz-Bauer, C., Aguzzi, A., 2008. Canine MDCK cell lines are refractory toinfection with human and mouse prions. Vaccine 26(21), 2601–-14.

\bibitem[Riek et al., 1996]{riek_etal1996}
Riek, R., Hornemann, S., Wider, G., Billeter, M., Glockshuber, R., Wüthrich, K., 1996. NMR structure of the mouse prion protein domain PrP(121--231). Nature 382(6587): 180--2.

\bibitem[Robakis et al., 1986]{robakis_etal1986}
Robakis, N.K., Sawh, P.R., Wolfe, G.C., Rubenstein, R., Carp, R.I., Innis, M.A., 1986. Isolation of a cDNA clone encoding the leader peptide of prion protein and expression of the homologous gene in various tissues. Proc. Natl. Acad. Sci. U. S. A. 83(17), 6377--81.

\bibitem[Roberts et al., 1988]{roberts_etal1988}
Roberts, G.W., Lofthouse, R., Allsop, D., Landon, M., Kidd, M., Prusiner, S.B., Crow, T.J., 1988. CNS amyloid proteins in neurodegenerative diseases. Neurology 38(10), 1534--40.

\bibitem[Rossetti et al., 2010]{rossetti_etal2010}
Rossetti, G., Giachin, G., Legname, G., Carloni, P., 2010. Structural facets of disease-linked human prion protein mutants: a molecular dynamic study. Proteins 78(16), 3270--80.

\bibitem[Sachsamanoglou et al., 2004]{sachsamanoglou_etal2004}
Sachsamanoglou, M., Paspaltsis, I., Petrakis, S., Verghese-Nikolakaki, S., Panagiotidis, C.H., Voigtlander, T., Budka, H., Langeveld, J.P., Sklaviadis, T., 2004. Antigenic profile of human recombinant PrP: generation and characterization of a versatile polyclonal antiserum. J. Neuroimmunol. 146(1--2), 22--32.

\bibitem[Sakudo et al., 2008]{sakudo_etal2008}
Sakudo, A., Nakamura, I., Tsuji, S., Ikuta, K., 2008. GPI-anchorless human prion protein is secreted and glycosylated but lacks superoxide dismutase activity. Int. J. Mol. Med. 21(2), 217--22.

\bibitem[Schmerr et al., 1994]{schmerr_etal1994}
Schmerr, M.J., Goodwin, K.R., Cutlip, R.C., 1994. Capillary electrophoresis of the scrapie prion protein from sheep brain. J. Chromatogr. A 680(2), 447--53.

\bibitem[Senator et al., 2004]{senator_etal2004}
Senator, A., Rachidi, W., Lehmann, S., Favier, A., Benboubetra, M., 2004. Prion protein protects against DNA damage induced by paraquat in cultured cells. Free Radic. Biol. Med. 37(8), 1224--30.

\bibitem[Shin et al., 2008]{shin_etal2008}
Shin, W., Lee, B., Hong, S., Ryou, C., Kwon, M., 2008. Cloning and expression of a prion protein (PrP) gene from Korean bovine (Bos taurus coreanae) and production of rabbit anti-bovine PrP antibody. Biotechnol. Lett. 30(10), 1705--11.

\bibitem[Shinagawa et al., 1986]{shinagawa_etal1986}
Shinagawa, M., Munekata, E., Doi, S., Takahashi, K., Goto, H., Sato, G., 1986. Immunoreactivity of a synthetic pentadecapeptide corresponding to the N-terminal region of the scrapie prion protein. J. Gen. Virol. 67 (Pt 8), 1745--50.

\bibitem[Sigurdson et al., 2011]{sigurdson_etal2011}
Sigurdson, C.J., Joshi-Barr, S., Bett, C., Winson, O., Manco, G., Schwarz, P., Rülicke, T., Nilsson, K.P., Margalith, I., Raeber, A., Peretz, D., Hornemann, S., Wüthrich, K., Aguzzi, A., 2011. Spongiform encephalopathy in transgenic mice expressing a point mutation in the $\beta$2-$\alpha$2 loop of the prion protein. J. Neurosci. 31(39), 13840--7.

\bibitem[Sigurdson et al., 2009]{sigurdson_etal2009}
Sigurdson, C.J., Nilsson, K.P., Hornemann, S., Heikenwalder, M., Manco, G., Schwarz, P., Ott, D., Rülicke, T., Liberski, P.P., Julius, C., Falsig, J., Stitz, L., Wüthrich, K., Aguzzi, A., 2009. De novo generation of a transmissible spongiform encephalopathy by mouse transgenesis. Proc. Natl. Acad. Sci. U. S. A. 106(1), 304--9.

\bibitem[Sigurdson et al., 2010]{sigurdson_etal2010}
Sigurdson, C.J., Nilsson, K.P., Hornemann, S., Manco, G., Fernández-Borges, N., Schwarz, P., Castilla, J., Wüthrich, K., Aguzzi, A., 2010. A molecular switch controls interspecies prion disease transmission in mice. J. Clin. Invest. 120(7), 2590--9.

\bibitem[Stanker et al., 2012]{stanker_etal2012}
Stanker, L.H., Scotcher, M.C., Lin, A., McGarvey, J., Prusiner, S.B., Hnasko, R., 2012. Novel epitopes identified by anti-PrP monoclonal antibodies produced following immunization of Prnp0/0 Balb/cJ mice with purified scrapie prions. Hybridoma. (Larchmt) 31(5), 314--24.

\bibitem[Sweeting et al., 2009]{sweeting_etal2009}
Sweeting, B., Brown, E., Chakrabartty, A., Pai, E.F., 2009. The structure of rabbit PrP$^\text{C}$: clues into species barrier and prion disease. Canadian Light Source 2009 Activity Report 28, pp. 72.

\bibitem[Sweeting et al., 2013]{sweeting_etal2013}
Sweeting, B., Brown, E., Khan, M.Q., Chakrabartty, A., Pai, E.F., 2013. N-terminal helix-cap in $\alpha$-helix 2 modulates $\beta$-state misfolding in rabbit and hamster prion proteins. PLoS One 8(5), e63047.
  
\bibitem[Takahashi et al., 1986]{takahashi_etal1986}
Takahashi, K., Shinagawa, M., Doi, S., Sasaki, S., Goto, H., Sato, G., 1986. Purification of scrapie agent from infected animal brains and raising of antibodies to the purified fraction. Microbiol. Immunol. 30(2), 123--31.

\bibitem[Takahashi et al., 1999]{takahashi_etal1999}
Takahashi, H., Takahashi, R.H., Hasegawa, H., Horiuchi, M., Shinagawa, M., Yokoyama, T., Kimura, K., Haritani, M., Kurata, T., Nagashima, K., 1999. Characterization of antibodies raised against bovine-PrP-peptides. J. Neurovirol. 5(3), 300--7.

\bibitem[Takekida et al., 2002]{takekida_etal2002}
Takekida, K., Kikuchi, Y., Yamazaki, T., Kakeya, T., Takatori, K., Tanamoto, K., Sawada, J., Tanimura, A., 2002. [Study on the detection of prion protein in food products by a competitive enzyme-linked immunosorbent assay]. Shokuhin. Eiseigaku. Zasshi. 43(3), 173--7.

\bibitem[Timmes et al., 2013]{timmes_etal2013}
Timmes, A.G., Moore, R.A., Fischer, E.R., Priola, S.A., 2013. Recombinant prion protein refolded with lipid and RNA has the biochemical hallmarks of a prion but lacks in vivo infectivity. PLoS One 8(7), e71081.

\bibitem[Tang et al., 2009]{tang_etal2009}
Tang, H., Fu, Y., Cui, Y., He, Y., Zeng, X., Ploplis, V.A., Castellino, F.J., Luo, Y., 2009. Fibrinogen has chaperone-like activity. Biochem. Biophys. Res. Commun. 378(3), 662--7.

\bibitem[Verdier, 2012]{verdier2012}
Verdier, J.M., 2012. Prions and Prion Diseases: New Developments, New York: NOVA Science Publishers, ISBN 978-1-62100-027-3, Preface, pp. vii-ix.

\bibitem[Vidal et al., 2013]{vidal_etal2013}
Vidal, E., Fernández-Borges, N., Pintado, B., Ordóñez, M., Márquez, M., Fondevila, D., Torres, J.M., Pumarola, M., Castilla, J., 2013. Bovine spongiform encephalopathy induces misfolding of alleged prion-resistant species cellular prion protein without altering its pathobiological features. J. Neurosci. 33(18), 7778--86.

\bibitem[Vilette et al., 2001]{vilette_etal2001}
Vilette, D., Andreoletti, O., Archer, F., Madelaine, M.F., Vilotte, J.L., Lehmann, S., Laude, H., 2001. Ex vivo propagation of infectious sheep scrapie agent in heterologous epithelial cells expressing ovine prion protein. Proc. Natl. Acad. Sci. U. S. A. 98, 4055--9.

\bibitem[Vol'pina et al., 2001]{volpina_etal2001}
Vol'pina, O.M., Zhmak, M.N., Obozhaia, M.B., Titova, M.A., Koroev, D.O., Volkova, T.D., Egorov, A.A., Ivanov, V.T., 2001. [Antibodies against synthetic fragments of the prion protein for the diagnosis of bovine spongiform encephalopathy]. Bioorg. Khim. 27(5), 352--8.

\bibitem[Vorberg et al., 2003]{vorberg_etal2003}
Vorberg, I., Martin, H.G., Eberhard, P., Suzette, A.P., 2003. Multiple amino acid residues within the rabbit prion protein inhibit formation of its abnormal isoform. J. Virol. 77, 2003--9.

\bibitem[Wade et al., 1987]{wade_etal1987}
Wade, W.F., Dees, C., German, T.L., Marsh, R.F., 1987. Immunochemical characterization of proteins from scrapie-infected hamster brain, using immunoblot analysis. Am. J. Vet. Res. 48(7), 1077--81.

\bibitem[Wang et al., 2013]{wangy_etal2013}
Wang, H., Yang, L., Kouadir, M., Tan, R., Wu, W., Zou, H., Wang, J., Khan, S.H., Li, D., Zhou, X., Yin, X., Wang, Y., Zhao, D., 2013. Expression and distribution of laminin receptor precursor/laminin receptor in rabbit tissues. J. Mol. Neurosci. 2013 May 30. [Epub ahead of print]

\bibitem[Wang et al., 2013]{wangz_etal2013}
Wang, Y., Zhao, S., Bai, L., Fan, J., Liu, E., 2013. Expression systems and species used for transgenic animal bioreactors. Biomed. Res. Int. 2013:580463. doi: 10.1155/2013/580463.

\bibitem[Wen et al., 2010a]{wen_etal2010a}
Wen, Y., Li, J., Xiong, M., Peng, Y., Yao, W., Hong, J., Lin, D., 2010. Solution structure and dynamics of the I214V mutant of the rabbit prion protein. PLoS One 5(10), e13273.

\bibitem[Wen et al., 2010b]{wen_etal2010b}
Wen, Y., Li, J., Yao, W., Xiong, M., Hong, J., Peng, Y., Xiao, G., Lin, D., 2010. Unique structural characteristics of the rabbit prion protein. J. Biol. Chem. 285(41), 31682--93.

\bibitem[Wiley et al., 1987]{wiley_etal1987}
Wiley, C.A., Burrola, P.G., Buchmeier, M.J., Wooddell, M.K., Barry, R.A., Prusiner, S.B., Lampert, P.W., 1987. Immuno-gold localization of prion filaments in scrapie-infected hamster brains. Lab. Invest. 57(6), 646--56.

\bibitem[Xi et al., 1994]{xi_etal1994}
Xi, Y.G., Cardone, F., Pocchiari, M., 1994. Detection of proteinase-resistant protein (PrP) in small brain tissue samples from Creutzfeldt-Jakob disease patients. J. Neurol. Sci. 124(2), 171--3.

\bibitem[Xiao et al., 2006]{xiao_etal2006}
Xiao, X.L., Han, J., Zhang, L., Chen, L., Zhang, J., Chen, X.L., Zhou, W., Jiang, H.Y., Zhang, B.Y., Liu, Y., Dong, X.P., 2006. Protein expression of human neuron-specific enolase and its antiserum preparation. Nan Fang Yi Ke Da Xue Xue Bao 26(11), 1543--7.

\bibitem[Yokoyama et al., 1996]{yokoyama_etal1996}
Yokoyama, T., Itohara, S., Yuasa, N., 1996. Detection of species specific epitopes of mouse and hamster prion proteins (PrPs) by anti-peptide antibodies. Arch. Virol. 141(3--4), 763--9.

\bibitem[Yokoyama et al., 1995]{yokoyama_etal1995}
Yokoyama, T., Kimura, K., Tagawa, Y., Yuasa, N., 1995. Preparation and characterization of antibodies against mouse prion protein (PrP) peptides. Clin. Diagn. Lab. Immunol. 2(2), 172--6.

\bibitem[Yost et al., 1990]{yost_etal1990}
Yost, C.S., Lopez, C.D., Prusiner, S.B., Myers, R.M., Lingappa, V.R., 1990. Non-hydrophobic extracytoplasmic determinant of stop transfer in the prion protein. Nature 343(6259), 669--72.

\bibitem[Zahn et al., 2000]{zahn_etal2000}
Zahn, R., Liu, A., Lührs, T., Riek, R., von Schroetter, C., López García, F., Billeter, M., Calzolai, L., Wider, G., Wüthrich, K., 2000. NMR solution structure of the human prion protein.  Proc. Natl. Acad. Sci. U. S. A. 97(1), 145--50.

\bibitem[Zhang, 2010]{zhang2010}
Zhang, J. 2010. Studies on the structural stability of rabbit prion protein probed by molecular dynamics simulations of its wild-type and mutants. J. Theor. Biol. 264, 119--22.

\bibitem[Zhang, 2011a] {zhang2011_horse}
Zhang, J., 2011. The structural stability of wild-type horse prion protein. J. Biomol. Struct. Dyn. 29(2), 369--77.

\bibitem[Zhang, 2011b]{zhang2011}
Zhang, J., 2011. Comparison studies of the structural stability of rabbit prion protein with human and mouse prion proteins. J. Theor. Biol. 269, 88--95.

\bibitem[Zhang, 2012]{zhang2012}
Zhang, J., 2012. The nature of the infectious agents: PrP models of resistant species to prion diseases (dog, rabbit and horses). In ``Prions and Prion Diseases: New Developments" (1st Edn), Verdier, J.M. (Ed), NOVA Science Publishers, New York, ISBN 978-1-61942-768-6: Chapt. 2, pp. 41--8.

\bibitem[Zhang and Liu, 2011]{zhang2011_dog}
Zhang, J., and Liu, D.D.W., 2011. Molecular dynamics studies on the structural stability of wild-type dog prion protein. J. Biomol. Struct. Dyn. 28(6), 861--9.

\bibitem[Zhao et al., 2000]{zhao_etal2000}
Zhao, X., Dong, X., Zhou, W., 2000. [Preparation of polyclonal antibody to human prion protein using the expressed GST-PrP fusion protein as antigen]. Zhonghua Shi Yan He Lin Chuang Bing Du Xue Za Zhi 14(2), 131--3.

\bibitem[Zhou, 1989]{zhou1989}
Zhou, G.P., 1989. Biological functions of soliton and extra electron motion in DNA structure. Phys. Scr. 40, 698--701.

\bibitem[Zhou, 2011a]{zhou2011a}
Zhou, G.P., 2011.  The disposition of the LZCC protein residues in wenxiang diagram provides new insights into the protein-protein interaction mechanism.  J. Theor. Biol. 284, 142--8.

\bibitem[Zhou, 2011b]{zhou2011b}
Zhou, G.P., 2011.  The structural determinations of the leucine zipper coiled-coil domains of the cGMP-dependent protein kinase I alpha and its interaction with the myosin binding subunit of the myosin light chains phosphase.  Proteins and Peptide Letters 18, 966--78.

\bibitem[Zhou and Huang, 2013]{zhouh2013}
Zhou, G.P. and Huang, R.B., 2013.  The pH-triggered conversion of the PrP$^{\text{C}}$ to PrP$^{\text{Sc}}$, Curr. Top. Med. Chem. 13(10), 1152--63.

\bibitem[Zhou et al., 2011]{zhou_etal2011}
Zhou, Z., Yan, X., Pan, K., Chen, J., Xie, Z.S., Xiao, G.F., Yang, F.Q., Liang, Y., 2011. Fibril formation of the rabbit/human/bovine prion proteins. Biophys. J .101(6), 1483--92.

\bibitem[Zocche et al., 2011]{zocche_etal2011}
Zocche Soprana, H., Canes Souza, L., Debbas, V., Martins Laurindo, F.R., 2011. Cellular prion protein (PrP$^{\text{C}}$) and superoxide dismutase (SOD) in vascular cells under oxidative stress. Exp. Toxicol. Pathol. 63(3), 229--36.
}
\end{thebibliography}

\end{document}